\documentclass[aps,reprint,superscriptaddress,nofootinbib]{revtex4-1}
\usepackage{amsmath}
\usepackage{esint}
\usepackage{amssymb}
\usepackage{stmaryrd}
\usepackage{subfigure}
\usepackage{psfrag}
\usepackage{graphicx}
\usepackage{color}
\usepackage[usenames,dvipsnames]{xcolor}
\usepackage{pifont}
\usepackage{fourier}
\usepackage{cancel}

\usepackage{epstopdf}
\usepackage[process=auto]{pstool}
\usepackage[normalem]{ulem}

\definecolor{red}{RGB}{237,28,36}
\definecolor{blue}{RGB}{57,83,164}
\definecolor{beige}{RGB}{250,164,26}
\definecolor{magenta}{RGB}{185,82,159}
\definecolor{green}{RGB}{0,161,75}
\definecolor{orange}{RGB}{241,90,41}

\newcommand{\figref}{Fig.~\ref}

\newcommand{\ve}[1]{\mathbf{#1}}
\newcommand{\ves}[1]{\boldsymbol{#1}}
\newcommand{\uve}[1]{\mathbf{\hat{{#1}}}}
\newcommand{\dya}[1]{\bar{\bar{#1}}}
\newcommand{\tx}[1]{\text{#1}}

\newcommand{\tr}[1]{{\cal T}\left\{#1\right\}}

\begin{document}

\title{What is Nonreciprocity?}

\author{Christophe Caloz}
\email[]{christophe.caloz@polymtl.ca}
\affiliation{Polytechnique Montr\'{e}al, Montr\'{e}al, QC H3T-1J4, Canada}
\author{Andrea Al\`{u}}
\affiliation{CUNY Advanced Science Research Center, NY 10031, USA, and The University of Texas at Austin, Austin, TX 78712, USA}
\author{Sergei Tretyakov}
\affiliation{Aalto University, Aalto, FI-00076, Finland}
\author{Dimitrios Sounas}
\affiliation{The University of Texas at Austin, Austin, TX 78712, USA}
\author{Karim Achouri}
\affiliation{EPFL, Lausanne, CH-1015, Switzerland}
\author{Zo\'{e}-Lise Deck-L\'{e}ger}
\affiliation{Polytechnique Montr\'{e}al, Montr\'{e}al, QC H3T-1J4, Canada}


\begin{abstract}
This paper aims at providing a global perspective on \emph{electromagnetic nonreciprocity} and clarifying confusions that arose in the recent developments of the field. It provides a general definition of nonreciprocity and classifies nonreciprocal systems according to their linear time-invariant (LTI), linear time-variant (LTV) or nonlinear nonreciprocal natures. The theory of nonlinear systems is established on the foundation of the concepts of time reversal, time-reversal symmetry, time-reversal symmetry breaking and related Onsager-Casimir relations. Special attention is given to LTI systems, as the most common nonreciprocal systems, for which a generalized form of the Lorentz reciprocity theorem is derived. The delicate issue of loss in nonreciprocal systems is demystified and the so-called thermodynamics paradox is resolved from energy conservation considerations. The fundamental characteristics and applications of LTI, LTV and nonlinear nonreciprocal systems are overviewed with the help of pedagogical examples. Finally, asymmetric structures with fallacious nonreciprocal appearances are debunked.
\end{abstract}

\maketitle

\section{Introduction}\label{sec:intro}
\emph{Nonreciprocity} arises in all branches of physics -- classical mechanics, thermodynamics and statistical mechanics, condensed matter physics, electromagnetism and electronics, optics, relativity, quantum mechanics, particle and nuclear physics, and cosmology -- underpinning a myriad of phenomena and applications.

In \emph{electromagnetics}, nonreciprocity is an important scientific and technological concept at both \emph{microwave}~\cite{Lax_1962,Pozar_ME_2011} and \emph{optical}~\cite{Born_1999,Saleh_Teich_FP_2007} frequencies\footnote{Historically, the development of \emph{commercial} nonreciprocal systems, following the scientific discovery of Faraday~\cite{Faraday_1933} and considerations of Rayleigh~\cite{Rayleigh_1901}, started in the microwave regime, following the invention of the magnetron cavity at the dawn of World War II, and experienced a peak in the period from 1950 to 1965~\cite{Lax_1962}. The development of nonreciprocal systems in the optical regime lagged its microwave counterpart by nearly 30 years, roughly corresponding to the time laps between the invention of the magnetron cavity and that of the laser.}. In both regimes, nonreciprocal devices have been almost exclusively based on ferrimagnetic (dielectric) compounds, called \emph{ferrites}, such as Yttrium Iron Garnet (YIG) and materials composed of iron oxides and other elements (Al, Co, Mn, Ni)~\cite{Gurevich_1996}.
Ferrite nonreciprocity results \emph{from electron spin precession\footnote{Landau-Lifshitz equation: $\partial\ve{m}/\partial t=-\gamma\ve{m}\times(\ve{B}_0+\mu_0\ve{H})$ ($\ve{m}$:~magnetic dipole moment; $\gamma$:~gyromagnetic ratio).} at microwaves}~\cite{Landau_1984,Schwinger_1998} and \emph{from electron cyclotron orbiting\footnote{Equation of motion for the electron: $(m_\tx{e}/e)\partial\ve{v}_\tx{e}/\partial t
=\ve{E}+\ve{v}_\tx{e}\times\ve{B}_0$ ($\ve{v}_\tx{e}$:~electron velocity; $e$:~electron charge; $m_\tx{e}$:~electron mass).} in optics}~\cite{Ishimaru_1990,Schwinger_1998}, with both effects being~\emph{induced by a static magnetic field bias $\ve{B}_0$}, which is provided by a permanent magnet or a resistive/superconductive coil.

However, ferrite-based systems tend to be bulky, heavy, costly, and non-amenable to integrated circuit technology, due to the incompatibility of ferrite crystal lattices with those of semiconductor materials. These issues have recently triggered an intensive quest for \emph{magnetless} nonreciprocity, i.e. nonreciprocity requiring no ferrimagnetic materials and magnets/coils.

This quest has led to the development of a wealth of \emph{magnetless nonreciprocal systems}, including metamaterials, space-time varying structures and nonlinear materials. However, it has also generated some confusion in the electromagnetics community~\cite{Fang_1996,Wang_OE_2011,Feng_Science_2011,Wang_SR_2012,Jalas_2013,Shi_2015,Fan_Science_2012}, related to the definition of nonreciprocity, the difference between linear and nonlinear nonreciprocity, the relation between nonreciprocity and time reversal symmetry breaking, the handling of time reversal in lossy and open systems, the ``thermodynamics paradox,'' and the difference between nonreciprocal and asymmetric propagation. The objective of this paper is to clarify this confusion, and provide a global perspective on electromagnetic nonreciprocity.

The paper is organized as follows. Section~\ref{sec:NR_def_class} defines and classifies nonreciprocal systems. Sections~\ref{sec:TRS}--\ref{sec:TRSB_example} explain the concepts of time reversal and time-reversal symmetry breaking, in general and specifically in electromagnetics.  Sections~\ref{sec:Lin_NR_media}--\ref{sec:gen_rec_th} study nonreciprocity in linear time-variant (LTI) media, culminating with the demonstration of a generalized form of the Lorentz theorem and the derivation of the Onsager-Casimir relations. Section~\ref{sec:Gen_Onsager_Casimir_rel} points out that these relations hold for all nonreciprocal systems, i.e. not only LTI but also linear time-variant (LTV) and nonlinear systems, while Sec.~\ref{sec:further_classif} specifies the applicability of the fundamental concepts established in the paper to these three types of nonreciprocal systems. Sections~\ref{sec:Lossy_syst} and~\ref{sec:open_syst} clarify the delicate handling of time-reversal symmetry in lossy and open systems, respectively. Based on the general definition of nonreciprocity, Sec.~\ref{sec:scat_pat_mod} extends the concept of S-parameters to all nonreciprocal systems. This serves as the foundation for fundamental energy conservation rules in Sec.~\ref{sec:energy_cons} and the resolution of the so-called thermodynamics paradox in Sec.~\ref{sec:thermo_par}. Building on previously established concepts, Secs.~\ref{sec:overview_three_NR_types}--\ref{sec:nonlin_syst} overview the fundamental characteristics and applications of LTI, LTV and nonlinear nonreciprocal systems. Finally, Sec.~\ref{sec:dist_asym} describes a few exotic systems whose asymmetries might be erroneously confused with nonreciprocity. Conclusions are given in Sec.~\ref{sec:concl} in the form of an enumeration of the main results of the paper.

\section{Nonreciprocity Definition and Classification}\label{sec:NR_def_class}

\emph{Nonreciprocity} is the absence of ``reciprocity.'' The adjective \emph{reciprocal} itself comes from the Latin word ``reciprocus''~\cite{Etymology_2001}, built on the prefixes \emph{re-} (backward) and \emph{pro-} (forward), that combine in the phrase \emph{reque proque} with the meaning of ``going backward as forward.'' Thus, ``reciprocal'' etymologically means ``going the same way backward as forward.''

In physics and engineering, the concept of nonreciprocity/reciprocity applies to \emph{systems}, which encompass \emph{media or structures} and \emph{components or devices}. \textbf{A \emph{nonreciprocal/reciprocal system} is defined as a system that exhibits different/same received-transmitted field ratios when its source(s) and detector(s) are exchanged}. In this definition, the notion of \emph{ratio} has been added to the aforementioned etymological meaning of ``reciprocity'' to reflect common practice.

\begin{table}[h]
\small
\centering
\begin{tabular}{@{\hskip -1mm}c|@{\hskip -1mm}c@{\hskip -1mm}c@{\hskip -0.5mm}}
&
 \begin{minipage}{0.21\textwidth}
 \centering
 \textbf{LINEAR} \\
 external bias TRS-B \\
 \underline{\textsc{strong nr form}}: \\
 arbitrary excitations \\
 arbitrary intensities \\
 high isolation
 \end{minipage}
&
 \begin{minipage}{0.21\textwidth}
 \centering
 \textbf{NONLINEAR (NL)} \\
 self bias $+$ str. asym. TRS-B \\
 \underline{\textsc{weak nr form}}: \\
 one excitation at a time \\
 intensity restrictions \\
 typically poor isolation
 \end{minipage}
\\RW0Q
\\[-0.8em]
\hline
\\[-0.8em]
\begin{minipage}{0.07\textwidth}\centering
\textsc{media} \\
or \textsc{struc-} \\
\textsc{tures}
\end{minipage}
&
\begin{minipage}{0.2\textwidth}\centering ferromagnets, ferrites \\ magnetized plasmas \\ 2DEGs, 2D materials \\ NR metamaterials \\ space-time media
\end{minipage}
&
\begin{minipage}{0.2\textwidth}\centering any strongly driven mat. \\ (e.g. glasses, crystals, semiconductors) \\ with spatial asymmetry, \\ NR metastructures
\end{minipage} \\
\\
\\[-1.6em]
\begin{minipage}{0.07\textwidth}\centering
\textsc{compo-} \\ \textsc{nents}  or \\ \textsc{devices}
\end{minipage}
&
\begin{minipage}{0.15\textwidth}\centering isolators \\ NR phase shifters \\ (e.g. gyrators) \\ circulators
\end{minipage}
&
\begin{minipage}{0.17\textwidth}\centering diodes \\ pseudo-isolators \\ power amplifiers \\ vacuum tubes
\end{minipage} \\
\end{tabular}
\caption{Classification and characteristics of nonreciprocal (NR) systems (TRS-B: Time-reversal symmetry breaking).}
\label{tab:classification}
\end{table}

As indicated in Tab.~\ref{tab:classification}, whose details will be discussed throughout the paper, nonreciprocal systems may be classified into \textbf{\emph{two fundamentally distinct categories}: \emph{linear} and \emph{nonlinear} nonreciprocal systems}~\cite{Potton_2004}. We shall see that in both cases nonreciprocity is based on \emph{time-reversal symmetry breaking}, by an \emph{external bias} in the linear case, and by a \emph{combination of self biasing and structural asymmetry} in the nonlinear case\footnote{Nonreciprocity based on externally-biased nonlinear media~\cite{Udalov_2012} has been little studied and has not led to practical implementation of nonreciprocal systems to date. Using an external bias, they would anyways loose their most interesting feature of being magnet-less or magnet-free.\label{fn:no_ext_bias_NL}}. We shall also see that linear nonreciprocity is stronger than nonlinear nonreciprocity, the former working for \emph{arbitrary excitations and intensities} with \emph{high isolation}, and the latter being restricted to \emph{nonsimultaneous excitations} from different directions, \emph{specific intensity conditions} and \emph{typically poor isolation}.

\section{Time Reversal and Time-Reversal Symmetry}\label{sec:TRS}

\begin{figure}[h]
    \centering
        \psfragfig*[width=\linewidth]{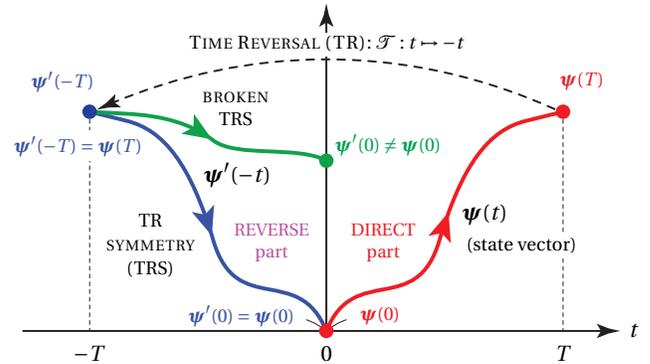}{
        \psfrag{t}[c][c]{$t$}
        \psfrag{0}[c][c]{$0$}
        \psfrag{T}[c][c]{$T$}
        \psfrag{M}[c][c]{$-T$}
        \psfrag{s}[c][c]{state}
        \psfrag{d}[c][c][0.85]{(state vector)}
        \psfrag{X}[l][l]{\textcolor{red}{$\ves{\psi}(t)$}}
        \psfrag{Y}[l][l]{\textcolor{blue}{$\ves{\psi}'(-t)$}}
        \psfrag{1}[l][l][0.85]{\textcolor{red}{$\ves{\psi}(0)$}}
        \psfrag{2}[c][c][0.85]{\textcolor{red}{$\ves{\psi}(T)$}}
        \psfrag{3}[c][c][0.85]{\textcolor{blue}{$\ves{\psi}'(-T)$}}
        \psfrag{x}[c][c][0.85]{\textcolor{blue}{$\ves{\psi}'(-T)=\ves{\psi}(T)$}}
        \psfrag{4}[l][l][0.85]{\textcolor{green}{$\ves{\psi}'(0)\neq \ves{\psi}(0)$}}
        \psfrag{5}[r][r][0.85]{\textcolor{blue}{$\ves{\psi}'(0)=\ves{\psi}(0)$}}
        \psfrag{B}[c][c][0.85]{\begin{minipage}{2cm}\centering TR \\ \textsc{symmetry} \\ (TRS)\end{minipage}}
        \psfrag{C}[c][c][0.85]{\begin{minipage}{2cm}\centering \textsc{broken} \\ TRS \end{minipage}}        \psfrag{D}[c][c]{$\ves{\psi}(t)$}
        \psfrag{E}[c][c]{$\ves{\psi}'(-t)$}
        \psfrag{F}[c][c][0.8]{\begin{minipage}{2cm}\centering\textcolor{red}{DIRECT} \\ \textcolor{red}{part}\end{minipage}}
        \psfrag{G}[c][c][0.8]{\begin{minipage}{2cm}\centering\textcolor{magenta}{REVERSE} \\  \textcolor{magenta}{part}\end{minipage}}
        \psfrag{R}[c][c][0.85]{\textsc{Time Reversal (TR):} ${\cal T}:t\mapsto-t$}}
        \caption{Time-reversal symmetry (TRS) (red and blue curves) and broken time-reversal symmetry (red and green curves) or time-reversal asymmetry as a general thought experiment and mathematical criterion for nonreciprocity in a 2-port system. Generalization to $N$-port systems or continuous media and nonlinear materials is straightforward upon implicitly considering reverse excitations at each port.}
   \label{fig:Time_Reversal}
\end{figure}

The etymological meaning of ``reciprocal'' as ``going the same way backward as forward'' suggests the \emph{thought experiment} depicted in Fig.~\ref{fig:Time_Reversal}.

In this experiment, one monitors a \emph{process} (temporal evolution of a physical phenomenon) occurring in a given \emph{system}, with \emph{ports} representing each a specific access point (or terminal), field mode and frequency range, \emph{as time is reversed}. Specifically, let us monitor the process between the ports P$_1$ and P$_2$ of the system via the state -- magnitude, phase, temporal frequency ($\omega$), spatial frequency ($\ve{k}$), polarization or spin angular momentum, orbital angular momentum -- vector $\ves{\Psi}(t)=[\psi_\tx{P$_1$}(t),\psi_\tx{P$_2$}(t)]^T$ ($T$: transpose), where $\psi_\tx{P$_{1,2}$}(t)$ represent the signal state at P$_{1,2}$. First, excite P$_1$ at $t=0$ and trace the response of the system until the time $t=T$ of complete transmission to P$_2$. Then, \emph{flip the sign of the time variable} in the process\footnote{While this operation is clearly unphysical (although causal), a corresponding physical operation may be envisioned via the shifting $\ves{\psi}'(-t)\rightarrow\ves{\psi}'(-t+T)$ or $\ves{\psi}'(-t)\rightarrow\ves{\psi}'(-t+2T+\Delta t)$, where the direct and reverse operations would be respectively simultaneous or successive with delay $\Delta t$.\label{footnote:shifting}}, which results in a system, not necessarily identical to the original one (Sec.~\ref{sec:TRS_breaking}), being excited at $t=-T$ and evolving until the time $t=0$ of complete transmission to P$_1$: this operation is called \textbf{\emph{Time Reversal}}, and it is a concept at the core of the works of Onsager in thermodynamics~\cite{Lewis_1925,Onsager_1931_I,Onsager_1931_II,Casimir_1945,Casimir_1963}\footnote{The central result of these works is the \emph{Onsager reciprocity relations}~\cite{Onsager_1931_I,Onsager_1931_II}, that led to the 1968 Nobel Prize in Physics and are sometimes dubbed the ``4$^\text{th}$ law of thermodynamics''~\cite{Wendt_1974}. These relations establish \emph{transfer function} equality relations (i.e. \emph{field ratio} equality relations, as the general definition of nonreciprocity in Sec.~\ref{sec:NR_def_class}), such as for instance between the heat flow per unit of pressure difference and the density flow per unit of temperature difference, as a consequence of time reversibility of microscopic dynamics. This is a general result, applying to all physical processes (e.g. transport of heat, electricity and matter) and even between different physical processes, as for instance the equality between the Peltier and Seebeck coefficients in thermoelectricity. \label{fn:Onsager}}.

Mathematically, time reversal is represented by the operator ${\cal T}$, defined as
\begin{equation}\label{eq:TR_op_def}
\tr{t}=t'=-t
\quad\text{or}\quad
{\cal T}:t\mapsto t'=-t
\end{equation}
when trivially applying to the time variable\footnote{This is meant as reversal of the \emph{time variable value} with \emph{fixed time coordinate direction}, i.e. \emph{symmetry w.r.t. axis $t=0$}, consistently with Fig.~\ref{fig:Time_Reversal}.}~\cite{Altman_2014},
and generally, when applying to a process, $\ves{\psi}(t)$, as
\begin{equation}\label{eq:TR_function}
\tr{\ves{\psi}(t)}=\ves{\psi}'(t')=\ves{\psi}'(-t).
\end{equation}

If the system remains the same/changes under time reversal, corresponding to the red-blue/green curve pairs in Fig.~\ref{fig:Time_Reversal}, i.e.
\begin{equation}\label{eq:TRS}
\tr{\ves{\psi}(t)}
=\ves{\psi}'(-t)
\begin{Bmatrix}
= \\
\neq
\end{Bmatrix}
\ves{\psi}(t),
\end{equation}
it is called \textbf{\emph{time-reversal symmetric/asymmetric}}. Since the direct and reverse parts of the process describe the response of the system for opposite transmission directions, time-reversal symmetry/asymmetry is intimately related to nonreciprocity/reciprocity. According to the definition in Sec.~\ref{sec:NR_def_class}, \emph{time-reversal symmetry/asymmetry} is \emph{equivalent to reciprocity/nonreciprocity} insofar as, in both cases, the system exhibits the/a same/different \emph{response} when transmitting at P$_1$ and receiving at P$_2$ as/than when transmitting at P$_2$ and receiving at P$_1$. Thus, \textbf{time-reversal symmetry/asymmetry provides a \emph{criterion for reciprocity/nonreciprocity}\footnote{This is a loose criterion since equivalence stricto senso requires equal direct and reverse \emph{field levels} and not just equal \emph{field ratios} (see Sec.~\ref{sec:Lossy_syst}).\label{fn:loose_def}}}.

\section{Time-Reversal Symmetry in Electromagnetics}\label{sec:TRS_EM}

\textbf{The basic \emph{laws of physics} are classically\footnote{The situation may different at the quantum level. Consider for instance two photons successively sent towards a dielectric slab from either side of it. Quantum probabilities may result in one photon being transmitted and the other reflected, i.e. a \emph{time-reversal asymmetric} process. However, repeating the experiment for more photons would invariably lead to time-reversal symmetric transmission-reflection ratios quickly converging to the classical scattering coefficients of the slab.} \emph{invariant  under time reversal, or time-reversal symmetric}}~\cite{Jackson_1998}, as may be intuitively understood by realizing that reversing time is equivalent to ``flipping the movie film'' of the process, as in Fig.~\ref{fig:Time_Reversal}. In contrast, the \emph{physical quantities} involved in the laws of physics, that we shall generically denote $f(t)$, may be either time-reversal symmetric or time-reversal antisymmetric, i.e.
\begin{equation}\label{eq:TR_eo}
\tr{f(t)}=f'(t')=f'(-t)=\pm f(-t),
\end{equation}
where ``$+$'' corresponds to \emph{time-reversal symmetry}, or \emph{even time-reversal parity}, and ``$-$'' corresponds to \emph{time-reversal antisymmetry}, or \emph{odd time-reversal parity}.

The \textbf{\emph{time-reversal parity of physical quantities}} may be easily inferred from fundamental laws. \textbf{Table~\ref{tab:Field_Symmetries}} presents the case of electromagnetic quantities~\cite{Jackson_1998}. Realizing that charges do not change as time passes, and are hence invariant under time reversal, allows one to sequentially deduce all the results of the figure by successively invoking Coulomb, Ohm, Amp\`{e}re, Maxwell, Poynting, impedance and Joule laws, equations and relations. Note that time reversal reverses the direction of wave propagation, consistently with the considerations in Sec.~\ref{sec:TRS}: $\tr{\ve{k}(t)}=-\ve{k}(-t)$.
\begin{table}[h]
    \centering
    \small
        \psfragfig*[width=0.97\linewidth]{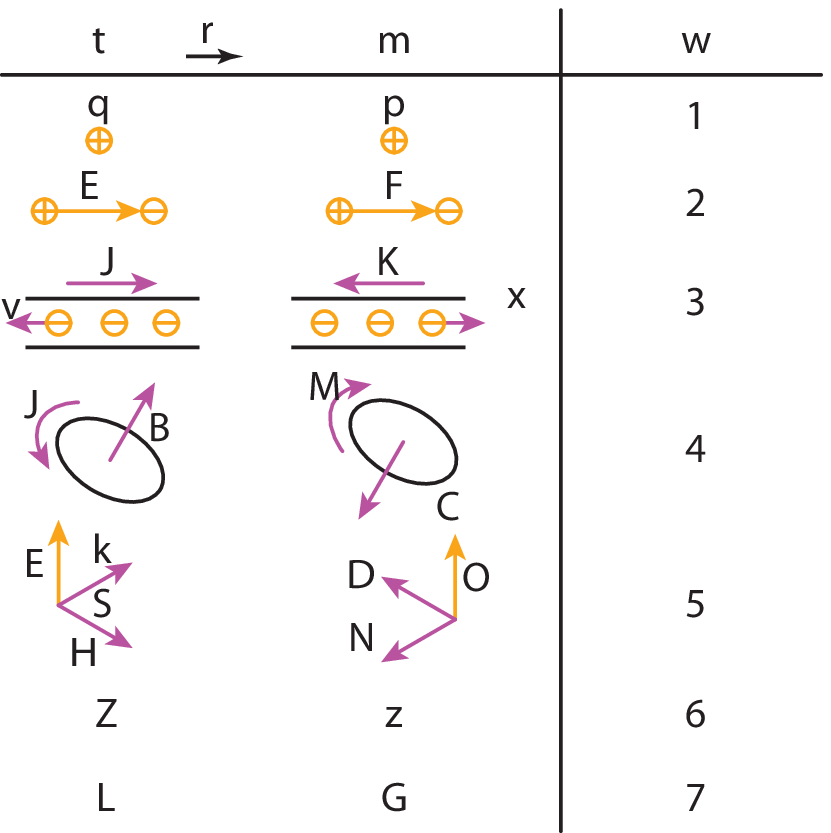}{
        \psfrag{t}[c][c]{$f(t)$}
        \psfrag{r}[c][c]{${\cal T}$}
        \psfrag{m}[c][c]{$f'(t')=\pm f(-t)$}
        \psfrag{w}[c][c]{\textsc{field -- parity}}
        \psfrag{1}[c][c]{\textcolor{beige}{$q,\rho$ -- even}}
        \psfrag{2}[c][c]{\textcolor{beige}{$\ve{E},\ve{P},\ve{D}$ -- even}}
        \psfrag{3}[c][c]{\textcolor{magenta}{$\ve{v},\ve{J}$ -- odd}}
        \psfrag{4}[c][c]{\textcolor{magenta}{$\ve{B},\ve{M},\ve{H}$ -- odd}}                        \psfrag{5}[cc][cc]{\textcolor{magenta}{$\ve{k},\ve{S}$ -- odd}}
        \psfrag{6}[cc][cc]{\textcolor{magenta}{$\eta$ -- odd}}
        \psfrag{7}[cc][cc]{\textcolor{magenta}{$\alpha$ -- odd}}
        \psfrag{q}[c][c][0.85]{\textcolor{beige}{$q$}}
        \psfrag{E}[c][c][0.85]{\textcolor{beige}{$\ve{E}$}}
        \psfrag{J}[c][c][0.85]{\textcolor{magenta}{$\ve{J}$}}
        \psfrag{v}[c][c][0.85]{\textcolor{magenta}{$\ve{v}$}}
        \psfrag{B}[c][c][0.85]{\textcolor{magenta}{$\ve{B}$}}
        \psfrag{H}[c][c][0.85]{\textcolor{magenta}{$\ve{H}$}}
        \psfrag{V}[c][c][0.85]{\textcolor{magenta}{$\ve{k},\ve{S}$}}        \psfrag{k}[l][l][0.85]{$\textcolor{magenta}{\ve{k}}=\textcolor{beige}{\ve{D}}\times\textcolor{magenta}{\ve{B}}$}
        \psfrag{S}[l][l][0.85]{$\textcolor{magenta}{\ve{S}}=\textcolor{beige}{\ve{E}}\times\textcolor{magenta}{\ve{H}}$}
        \psfrag{p}[c][c][0.85]{\textcolor{beige}{$q'=q$}}
        \psfrag{F}[c][c][0.85]{\textcolor{beige}{$\ve{E}'=\ve{E}$}}
        \psfrag{K}[c][c][0.85]{\textcolor{magenta}{$\ve{J}'=-\ve{J}$}}
        \psfrag{x}[ct][ct][0.85]{\begin{minipage}{1cm}\textcolor{magenta}{$\ve{v}'=$ \\ $\phantom{..}-\ve{v}$}\end{minipage}}
        \psfrag{M}[c][c][0.85]{\textcolor{magenta}{$\ve{J}'$}}
        \psfrag{C}[c][c][0.85]{\textcolor{magenta}{$\ve{B}'=-\ve{B}$}}
        \psfrag{D}[c][c][0.85]{\textcolor{magenta}{$\ve{H}'$}}
        \psfrag{I}[c][c][0.85]{\textcolor{magenta}{$\ve{E}'$}}
        \psfrag{N}[c][c][0.85]{\textcolor{magenta}{$\ve{k}',\ve{S}'$}}
        \psfrag{O}[c][c][0.85]{\textcolor{beige}{$\ve{E}'$}}         \psfrag{k}[l][l][0.85]{$\textcolor{magenta}{\ve{k}}=\textcolor{beige}{\ve{D}}\times\textcolor{magenta}{\ve{B}}$}
        \psfrag{S}[l][l][0.85]{$\textcolor{magenta}{\ve{S}}=\textcolor{beige}{\ve{E}}\times\textcolor{magenta}{\ve{H}}$}
        \psfrag{Z}[c][c][0.85]{$\textcolor{magenta}{\eta}=\textcolor{beige}{E}/\textcolor{magenta}{H}$}
        \psfrag{z}[c][c][0.85]{$\textcolor{magenta}{\eta'}=\textcolor{beige}{E'}/\textcolor{magenta}{H'}=\textcolor{magenta}{-\eta}$}
        \psfrag{L}[c][c][0.85]{\begin{minipage}{3.4cm}\centering
            $\textcolor{magenta}{\alpha}=\Im\{\textcolor{magenta}{k}\}<0$ \\ loss \end{minipage}}
        \psfrag{G}[c][c][0.85]{\begin{minipage}{3.4cm}\centering
            $\textcolor{magenta}{\alpha'}=\Im\{\textcolor{magenta}{k'}\}=\textcolor{magenta}{-\alpha}>0$ \\ gain \end{minipage}}
        }
        \caption{Time-reversal parity (even/odd), or symmetry/antisymmetry, of the main electromagnetic quantities (time-harmonic dependence $e^{j\omega t}$).}
   \label{tab:Field_Symmetries}
\end{table}

The \textbf{\emph{time-reversal symmetry of Maxwell equations}} may be straightforwardly verified from applying the time-reversal rules in Tab.~\ref{tab:Field_Symmetries} and noting that the $\nabla$ operator is time-reversal invariant. Specifically, replacing all the primed quantities in time-reversed Maxwell equations from their unprimed (original) counterparts according to Tab.~\ref{tab:Field_Symmetries}, and simplifying signs, restores the original Maxwell equations, i.e.
\begin{subequations}\label{eq:Maxwell}
\begin{equation}\label{eq:Maxwell_Faraday}
\nabla\times\ve{E}^{(\prime)}=-\partial\ve{B}^{(\prime)}/\partial t^{(\prime)}
\;\left(\text{even}\equiv\text{odd}/\text{odd}\right),
\end{equation}
\begin{equation}\label{eq:Maxwell_Ampere}
\nabla\times\ve{H}^{(\prime)}=\partial\ve{D}^{(\prime)}/\partial t^{(\prime)}+\ve{J^{(\prime)}}
\;\left(\text{odd}\equiv\text{even}/\text{odd}+\text{odd}\right),
\end{equation}
\end{subequations}
with parity matching indicated in brackets~\cite{Supp_Mat_NR}. The time-reversal invariance of Maxwell equations indicates that if \emph{\underline{all} the quantities of an electromagnetic system are time-reversed}, according to the time-reversal parity rules in Tab.~\ref{tab:Field_Symmetries}, \emph{then the time-reversal system will have the same electromagnetic solution as the direct system, whatever its complexity}!

\section{Time-Reversal Symmetry Breaking \\ and Nonreciprocity}\label{sec:TRS_breaking}

\textbf{\emph{Time-reversal symmetry breaking} is an operation that destroys, with an external or internal (due to wave itself) \emph{bias}, the time symmetry of a process, and hence makes it time-reversal asymmetric, by violating (at least) one of the time-reversal rules}, such as those of Tab.~\ref{tab:Field_Symmetries}. Since all physical quantities are either even or odd under time reversal [Eq.~\eqref{eq:TR_eo}], \emph{Time-reversal symmetry breaking requires reversing/maintaining the sign of \underline{at least one} of the time-reversal even/odd quantities (bias) involved in the system}. Only the latter option, viz. maintaining the sign of a time-reversal odd quantity, such as for instance $\ve{v}$, $\ve{J}$ or $\ve{B}$ in Tab.~\ref{tab:classification}, is practically meaningful, since this is the only case which leaves the system the same. This fact will become more clear in Sec.~\ref{sec:gen_rec_th}.

The \textbf{time-reversal symmetry/asymmetry criterion} for determining the non/reciprocity of a given system (Sec.~\ref{sec:TRS}) may be applied as follows. First, \textbf{one fully time-reverses the system using the rules in Tab.~\ref{tab:Field_Symmetries}}. As a result, the process retrieves its initial state. However, \emph{the time-reversal operation may have altered the nature of the system}, resulting in \emph{different direct and reverse systems}. In such a case, the time-reversal experiment is irrelevant, comparing apples and pears. So one must \textbf{examine whether the reversed system is identical to the given one or not}. If it is, the process is time-reversal symmetric and the system is reciprocal. Otherwise, the system must violate a time-reversal rule to maintain its nature, or break time-reversal symmetry (or become time-reversal asymmetric), and is hence nonreciprocal. This is assuming no loss/gain. The case of loss/gain (last row in the Tab.~\ref{tab:Field_Symmetries}) requires special attention and will be treated in Secs.~\ref{sec:Lossy_syst} and~\ref{sec:open_syst}.

\section{Electromagnetic Example}\label{sec:TRSB_example}

\begin{figure}[h]
    \centering
        \psfragfig*[width=\linewidth]{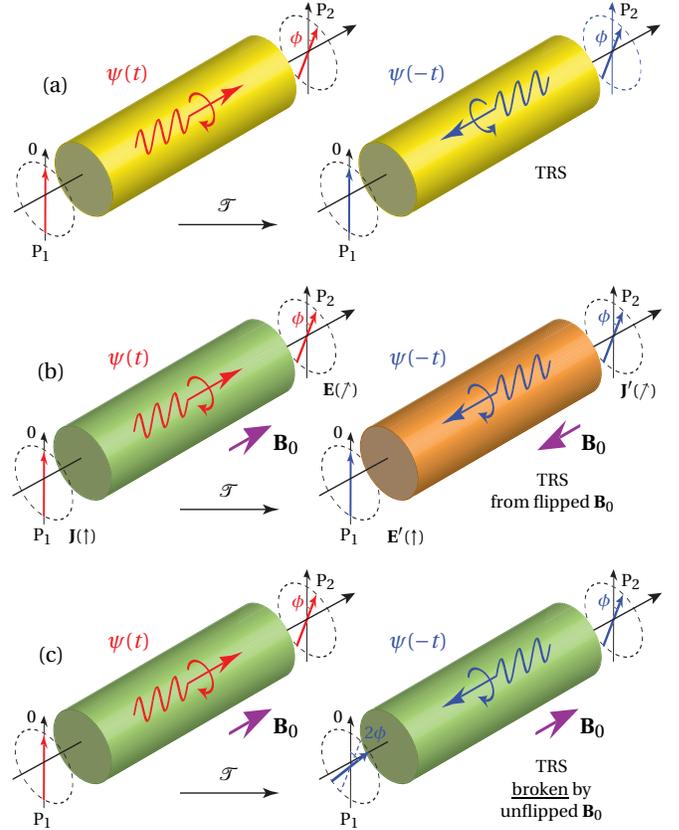}{
        \psfrag{A}[c][c]{(a)}
        \psfrag{B}[c][c]{(b)}
        \psfrag{C}[c][c]{(c)}
        \psfrag{K}[c][c][0.8]{$\ve{J}(\uparrow)$}
        \psfrag{L}[c][c][0.8]{$\ve{E}(\nnearrow)$}
        \psfrag{M}[c][c][0.8]{$\ve{E}'(\uparrow)$}
        \psfrag{N}[c][c][0.8]{$\ve{J}'(\nnearrow)$}
        \psfrag{0}[c][c][0.8]{0}
        \psfrag{1}[c][c][0.8]{P$_1$}
        \psfrag{2}[c][c][0.8]{P$_2$}
        \psfrag{H}[c][c]{$\ve{B}_0$}
        \psfrag{X}[l][l]{\textcolor{red}{$\psi(t)$}}
        \psfrag{Y}[l][l]{\textcolor{blue}{$\psi(-t)$}}
        \psfrag{p}[c][c][0.8]{\textcolor{red}{$\phi$}}
        \psfrag{r}[c][c][0.8]{\textcolor{blue}{$\phi$}}
        \psfrag{q}[c][c][0.8]{\textcolor{blue}{$2\phi$}}
        \psfrag{t}[c][c][0.8]{${\cal T}$}
        \psfrag{a}[c][c][0.8]{TRS}
        \psfrag{b}[c][c][0.8]{\begin{minipage}{4cm}\centering
            TRS \\ from flipped $\ve{B}_0$\end{minipage}}
        \psfrag{c}[c][c][0.8]{\begin{minipage}{3cm}\centering
            TRS \\ \underline{broken} by \\unflipped $\ve{B}_0$\end{minipage}}
        }
        \caption{Application of the time-reversal asymmetry/symmetry criterion for non/reciprocity (Sec.~\ref{sec:TRS_breaking}) to two systems inducing electromagnetic field polarization rotation (process $\equiv$ gyrotropy). (a)~Chiral system, without bias, and hence time-reversal symmetric (TRS) [$\psi_{\tx{P}_1}(0)=E^{\text{P}_1}=E_0$ $\mapsto$ $\psi'_{\tx{P}_1}(0)=E_0=\psi(0)$], i.e. reciprocal. (b)~Faraday system \emph{altered by time-reversal odd bias $\ve{B}_0$ flipped according to time reversal} (irrelevant time-reversal test due to altered nature of the system). (c)~Same as (b) but with \emph{unflipped} $\ve{B}_0$, and hence unaltered system, breaking the time-reversal symmetry of the process [$\psi_{\tx{P}_1}'(0)=E_0\cos(2\phi)\neq E_0=\psi_{\tx{P}_1}(0)$] and hence revealing nonreciprocity.}
   \label{fig:Gyro_Rec_Med}
\end{figure}

To better grasp the concepts of Secs.~\ref{sec:TRS}, \ref{sec:TRS_EM} and \ref{sec:TRS_breaking}, consider Fig.~\ref{fig:Gyro_Rec_Med}, involving two gyrotropic systems, a chiral system~\cite{Arago_1812,Pasteur_1850,Bose_1898,Lindman_1914,Sihvola_1994,Kong_2008} and a Faraday system~\cite{Crowther_1920,Faraday_1933,Budden_1954,Gurevich_1996,Pozar_ME_2011}.

In the case of the \emph{chiral system}, in Fig.~\ref{fig:Gyro_Rec_Med}(a), the field polarization is rotated along the chiral medium according to the \emph{handedness of the helix-shaped particles} that compose it. Upon time reversal, the direction of propagation is reversed, according to Tab.~\ref{tab:Field_Symmetries}. Thus, the field polarization symmetrically returns to its original state, as the current rewinds along the particles, \emph{without any system alteration}. Chiral (unbiased) gyrotropy is thus a time-reversal symmetric process, and is therefore reciprocal.

Now consider the \emph{Faraday system}, in Figs.~\ref{fig:Gyro_Rec_Med}(b) and (c). In such a system, the direction of polarization rotation is not any more dictated by particle shape, but by a \emph{static magnetic field, $\ve{B}_0$, or bias,} provided by an external magnet and inducing specific spin states in the medium at the atomic level.

Full time reversal requires here reversing the sign of $\ve{B}_0$, as in Fig.~\ref{fig:Gyro_Rec_Med}(b), just as for any other time-reversal odd quantity. Then, waves propagating in opposite directions indeed see the same effective medium by symmetry. However, the system has been \emph{altered} upon time reversal, since its spins have been reversed. Therefore, the time-reversal experiment is irrelevant to nonreciprocity! In order to preserve the nature of this system, and hence properly decide on its (non)reciprocity, one must preserve its spin states by keeping the direction of $\ve{B}_0$ unchanged, as shown in Fig.~\ref{fig:Gyro_Rec_Med}(c) and done in practice. But this violates a time-reversal symmetry rule, i.e. breaks time-reversal symmetry, or makes the process time-reversal asymmetric, revealing ipso facto that the \emph{Faraday system is nonreciprocal}.

\section{Linear Nonreciprocal Media}\label{sec:Lin_NR_media}

The vast majority of current nonreciprocal systems are based on \emph{LTI media} or, more precisely, media whose nonreciprocity is enabled by an external bias rather than nonreciprocity with structural asymmetry (Tab.~\ref{tab:classification}). We shall therefore dedicate the remainder of this section and Secs.~\ref{sec:ser_rec_from_TRS}--\ref{sec:gen_rec_th} to the study of nonreciprocity in such media, and will later generalize the discussion to the case of the linear time-variant (LTV) and nonlinear nonreciprocal systems.

The example of Sec.~\ref{sec:TRSB_example} has illustrated how nonreciprocity is achieved by breaking time-reversal symmetry with an \emph{external} \emph{bias}, that is a magnetic field. Alternative time-reversal odd biases are the velocity or the current (Tab.~\ref{tab:Field_Symmetries}), and we therefore generically denote the bias field $\ve{F}_0$.

Linear nonreciprocal media (Tab.~\ref{tab:classification}) include a)~\emph{ferromagnets} (magnetic conductors) and \emph{ferrites}~\cite{Smit_1958}, whose nonreciprocity is based on electron spin precession (permeability tensor with same structure)~\cite{Gurevich_1996,Lax_1962,Rodrigue_1988} or cyclotron orbiting (permittivity tensor)~\cite{Landau_1984,Zvezdin_Kotov_1997,Jackson_1998}, due to a magnetic field bias, b)~\emph{magnetized plasmas}~\cite{Ishimaru_1990,Stix_1992}, \emph{two-dimensional electron gases} (e.g. GaAs, GaN, InP)~\cite{Chiu_1974,Cavalli_1993,Allis_2003,Chin_2013} and other \emph{2D materials}, such as graphene~\cite{Sounas_APL_01_2011,Crassee_2011,Sounas_TMTT_04_2012,Chamanara_MWCL_07_2012,Chamanara_OE_05_2013,Sounas_APL_05_2013,Tamagnone_2014,Tamagnone_2016}, whose nonreciprocity is again based on cyclotron orbiting due to a magnetic field bias (permittivity tensor), c)~\emph{space-time moving/modulated media}, whose nonreciprocity is based on the motion of matter/propagation of a perturbation associated with a force/current bias~\cite{Cassedy_1963,Cassedy_1965,Cassedy_1967,Chu_1972,Kippenberg_2007,Yu_2009,Kang_2011,Estep_2014,Taravati_TAP_02_2017,Taravati_PRB_10_2017,Chamanara_PRB_10_2017}, and d)~\emph{transistor-loaded metamaterials}, mimicking ferrites~\cite{Kodera_APL_07_2011,Kodera_AWPL_01_2012,Kodera_AWPL_12_2012,Sounas_TAP_01_2013,Kodera_2018} or using twisted dipoles~\cite{Wang_PNAS_2012}, both based on a current bias.

The \emph{constitutive relations} of media are ideally expressed in the frequency domain, since molecules act as small oscillators with specific resonances. In the case of a \textbf{\emph{linear time-invariant (LTI)\footnote{The LTV counterpart of~\eqref{eq:TR_bianisotropic_rel} would involve convolution products.} bianisotropic medium}}~\cite{Kong_1972,Kong_2008,Rothwell_2008}, these relations may be written, for the given medium and its time-reversed counterpart, as
\begin{subequations}\label{eq:TR_bianisotropic_rel}
\begin{equation}\label{eq:TR_bianisotropic_rel_D}
\tilde{\ve{D}}^{(\prime)}
=\tilde{\dya{\epsilon}}^{(\prime)}(\pm\ve{F}_0)\cdot\tilde{\ve{E}}^{(\prime)}
+\tilde{\dya{\xi}}^{(\prime)}(\pm\ve{F}_0)\cdot\tilde{\ve{H}}^{(\prime)},
\end{equation}
\begin{equation}\label{eq:TR_bianisotropic_rel_B}
\tilde{\ve{B}}^{(\prime)}
=\tilde{\dya{\zeta}}^{(\prime)}(\pm\ve{F}_0)\cdot\tilde{\ve{E}}^{(\prime)}
+\tilde{\dya{\mu}}^{(\prime)}(\pm\ve{F}_0)\cdot\tilde{\ve{H}}^{(\prime)},
\end{equation}
\end{subequations}
where the temporal frequency ($\omega$) dependence is implicitly assumed everywhere, where $\tilde{\dya{\epsilon}}$, $\tilde{\dya{\mu}}$, $\tilde{\dya{\xi}}$ and $\tilde{\dya{\zeta}}$ are the frequency-domain medium permittivity, permeability, magneto-electric coupling and electro-magnetic coupling complex dyadic functions, respectively, and where the ``$+$'' and ``$-$'' signs of the (time-reversal odd) bias $\ve{F}_0$ correspond to the unprimed (given) and primed (time-reversal symmetric) problems, respectively. For instance, in a ferrite at microwaves\footnote{In optics, it is the permittivity that is tensorial (magneto-optic effect~\cite{Zvezdin_Kotov_1997}).}, $\tilde{\dya{\xi}}=\tilde{\dya{\zeta}}=0$, $\tilde{\dya{\epsilon}}=\epsilon\dya{I}$ and, if $\ve{B}_0\|\uve{z}$, $\tilde{\dya{\mu}}(\omega)=\mu_\tx{d}\dya{I}_t+j\mu_\tx{o}\dya{J}+\mu_0\uve{z}\uve{z}$ ($\dya{I}$: unit dyadic, $\dya{I}_t=\dya{I}-\uve{z}\uve{z}$, $\dya{J}=\dya{I}_t\times\uve{z}=\uve{x}\uve{y}-\uve{y}\uve{x}$), where $\mu_\tx{d}=\mu_0[1+\omega_0\omega_\tx{m}/(\omega_0^2-\omega^2)]$ and $\mu_\tx{o}=\mu_0\omega\omega_\tx{m}/(\omega_0^2-\omega^2)$, with $\omega_0=\gamma B_0$ (ferrimagnetic resonance, $\gamma$: gyromagnetic ratio) and $\omega_\tx{m}=\mu_0\gamma M_\tx{s}$ ($M_\tx{s}$: saturation magnetization)\footnote{If $\ve{B}_0=0$, $\tilde{\dya{\mu}}(\omega)=\mu_0I$ (ferrite purely dielectric, $\epsilon_\tx{r}\approx 10-15$~\cite{Lax_1962}).}, which extends to the lossy case using $\omega_0\rightarrow\omega_0+j\omega\alpha$ ($\alpha$: damping factor)~\cite{Lax_1962,Gurevich_1996,Pozar_ME_2011}.

\section{Time Reversal in the Frequency Domain}\label{sec:ser_rec_from_TRS}

How \emph{frequency-domain fields} transform under time reversal may be found via Fourier transformation with time-reversal rules (Sec.~\ref{sec:TRS_EM}) as
\begin{equation}\label{eq:TR_Pw}
\begin{split}
    \tr{\tilde{f}(\omega)}
    &=\tilde{f}'(\omega)=\tr{\int_{-\infty}^{+\infty}f(t)e^{-j\omega t}dt} \\
    &=\int_{+\infty}^{-\infty}\left[\pm f(-t)\right\}e^{+j\omega t}(-dt)
    \overset{\text{Eq.~\eqref{eq:TR_eo}}}{=}\pm \tilde{f}^*(\omega),
\end{split}
\end{equation}
where the ``$+$'' and ``$-$'' signs correspond to time-reversal even and time-reversal odd quantities, respectively~(Tab.~\ref{tab:Field_Symmetries}).

One may next infer from Eq.~\eqref{eq:TR_Pw} how the \emph{frequency-domain media parameters} transform under time reversal. For this purpose, compare the given (unprimed) medium in~\eqref{eq:TR_bianisotropic_rel} and its time-reversed (primed) counterpart with time-reversed field substitutions~\eqref{eq:TR_Pw}. This yields~\cite{Supp_Mat_NR}
\begin{subequations}\label{eq:freq_const_rel}
\begin{equation}
\tilde{\dya{\epsilon}}'(\ve{F}_0)
=\tilde{\dya{\epsilon}}^*(-\ve{F}_0),
\quad
\tilde{\dya{\mu}}'(\ve{F}_0)
=\tilde{\dya{\mu}}^*(-\ve{F}_0),
\end{equation}
\begin{equation}
\tilde{\dya{\xi}}'(\ve{F}_0)
=-\tilde{\dya{\xi}}^*(-\ve{F}_0),
\quad
\tilde{\dya{\zeta}}'(\ve{F}_0)
=-\tilde{\dya{\zeta}}^*(-\ve{F}_0).
\end{equation}
\end{subequations}
Thus, \textbf{frequency-domain time reversal implies \emph{complex conjugation} plus \emph{proper parity transformation}}. One may easily verify that inserting~\eqref{eq:TR_Pw} and~\eqref{eq:freq_const_rel} into~\eqref{eq:Maxwell} and~\eqref{eq:TR_bianisotropic_rel} transforms the \emph{time-reversed Maxwell and constitutive equations} to \emph{equations that are exactly identical to their respective original forms, except for the substitution of all the quantities by their complex conjugate.} Time reversal without `$*$' in~\eqref{eq:freq_const_rel}, i.e. not transforming loss into gain, is called \emph{restricted time reversal}~\cite{Altman_2014}, and will be used in the next section to derive the generalized Lorentz (non)reciprocity theorem for LTI systems.

\section{Generalized Lorentz Reciprocity Theorem \\ and Onsager-Casimir Relations}\label{sec:gen_rec_th}

Applying the usual reciprocity manipulations of Maxwell equations~\cite{Lorentz_1896,Sommerfeld_1925,Kong_2008,Rothwell_2008} to the frequency-domain version of~\eqref{eq:Maxwell} with the time-reversal transformations~\eqref{eq:TR_Pw} yields~\cite{Supp_Mat_NR}
%
\begin{gather}
\iiint_{V_J}\tilde{\ve{J}}\cdot\tilde{\ve{E}}^*dv
-\iiint_{V_{J}}\tilde{\ve{J}}^*\cdot\tilde{\ve{E}}dv
=\oiint_S\left(\tilde{\ve{E}}\times\tilde{\ve{H}}^*
-\tilde{\ve{E}}^*\times\tilde{\ve{H}}\right)\cdot\uve{n}ds \nonumber \\
-j\omega\iiint_V\left(\tilde{\ve{E}}^*\cdot\tilde{\ve{D}}
-\tilde{\ve{E}}\cdot\tilde{\ve{D}}^*+\tilde{\ve{H}}\cdot\tilde{\ve{B}}^*
-\tilde{\ve{H}}^*\cdot\tilde{\ve{B}}\right)dv. \nonumber \\
\label{eq:reac_EDHB}
\end{gather}
If the medium is unbounded, so that $[\uve{n}\times\ve{E}^{(*)}=\eta\ve{H}^{(*)}]_S$ assuming \emph{restricted time reversal} ($\eta$ unchanged), or enclosed by an impenetrable cavity, the surface integral in this equation vanishes~\cite{Supp_Mat_NR}. In reciprocal systems, the LHS (reaction difference~\cite{Kong_2008}) also vanishes, as found by first applying Eq.~\eqref{eq:reac_EDHB} to vacuum, as a fundamental reciprocity condition in terms of system ports. Inserting~\eqref{eq:TR_bianisotropic_rel}, transformed according to the \emph{restricted time reversal} version of~\eqref{eq:freq_const_rel} (no `$*$'), in the resulting relation yields~\cite{Supp_Mat_NR}
\begin{gather}
\iiint_V\Big\{
\tilde{\ve{E}}^*\cdot\left[\tilde{\dya{\epsilon}}(\ve{F}_0)
-\tilde{\dya{\epsilon}}^T(-\ve{F}_0)\right]\cdot\tilde{\ve{E}}
-\tilde{\ve{H}}^*\cdot\left[\tilde{\dya{\mu}}(\ve{F}_0)\right. \nonumber \\
-\left.\tilde{\dya{\mu}}^T(-\ve{F}_0)\right]\cdot\tilde{\ve{H}}
+\tilde{\ve{E}}^*\cdot\left[\tilde{\dya{\xi}}(\ve{F}_0)
+\tilde{\dya{\zeta}}^T(-\ve{F}_0)\right]\cdot\tilde{\ve{H}} \nonumber  \\
-\tilde{\ve{H}}^*\cdot\left[\tilde{\dya{\zeta}}(\ve{F}_0)
+\tilde{\dya{\xi}}^T(-\ve{F}_0)\right]\cdot\tilde{\ve{E}}\Big\}dv
=0,\label{eq:gen_Lorentz}
\end{gather}
where the identity $\ve{a}\cdot\dya{\chi}\cdot\ve{b}
=\left(\ve{a}\cdot\dya{\chi}\cdot\ve{b}
\right)^T=\ve{b}\cdot\dya{\chi}^T\cdot\ve{a}$ ($T$: transpose), has been used to group dyadics with opposite pre-/post-multiplying fields. Equation~\eqref{eq:gen_Lorentz} represents the \textbf{\emph{generalized form of the Lorentz reciprocity theorem for LTI bianisotropic media}}~\cite{Casimir_1945,Serdyukov_2001}. Since this relation must hold for arbitrary fields, one must have
\begin{subequations}\label{eq:Onsager_Casimir_rel}
\begin{equation}\label{eq:Onsager_Casimir_rel_eps}
\tilde{\dya{\epsilon}}(\ve{F}_0)
=\tilde{\dya{\epsilon}}^T(-\ve{F}_0),
\end{equation}
\begin{equation}\label{eq:Onsager_Casimir_rel_mu}
\tilde{\dya{\mu}}(\ve{F}_0)=\tilde{\dya{\mu}}^T(-\ve{F}_0),
\end{equation}
\begin{equation}\label{eq:Onsager_Casimir_rel_chi_zeta}
\tilde{\dya{\xi}}(\ve{F}_0)
=-\tilde{\dya{\zeta}}^T(-\ve{F}_0).
\end{equation}
\end{subequations}
which are the electromagnetic version of the
\textbf{\emph{Onsager-Casimir reciprocity relations}}\footnote{If the medium is \emph{lossless}, we also have $\tilde{\dya{\epsilon}}=\tilde{\dya{\epsilon}}^\dagger$, $\tilde{\dya{\mu}}=\tilde{\dya{\mu}}^\dagger$, $\tilde{\dya{\xi}}=\tilde{\dya{\zeta}}^\dagger$~\cite{Kong_2008}, leading to the additional constraint $\Im\{\tilde{\dya{\epsilon}}\}=\Im\{\tilde{\dya{\mu}}\}=\Re\{\tilde{\dya{\xi}}\}=\Re\{\tilde{\dya{\zeta}}\}=0$.}$^,$, with negation corresponding to nonreciprocity. If $\ve{F}_0=0$ (no bias), Eqs.~\eqref{eq:Onsager_Casimir_rel} reduce to the conventional reciprocity relations $\tilde{\dya{\epsilon}}=\tilde{\dya{\epsilon}}^T$, $\tilde{\dya{\mu}}=\tilde{\dya{\mu}}^T$ and $\tilde{\dya{\xi}}=-\tilde{\dya{\zeta}}^T$~\cite{Kong_2008}.

It may be shown~\cite{Supp_Mat_NR} that the transposed dyadics in~\eqref{eq:Onsager_Casimir_rel} correspond to propagation in the reverse direction than that of the wave associated with the untransposed dyadics, as they stem from the conjugate quantities $\tilde{\ve{D}}^*$ and $\tilde{\ve{B}}^*$ in~\eqref{eq:reac_EDHB} and as conjugate fields (with proper parity) are equivalent to reversed fields [Eq.~\ref{eq:TR_Pw}]. Thus, the Onsager-Casimir relations~\eqref{eq:Onsager_Casimir_rel} \emph{reveal} that \textbf{nonreciprocity results from not reversing $\ve{F}_0$ so that waves propagating in opposite directions see different effective media}\footnote{See the example of Figs.~\ref{fig:Gyro_Rec_Med}(b) and (c) with the ferrite $\tilde{\dya{\mu}}$ tensor in Sec.~\ref{sec:Lin_NR_media}.}$^,$\footnote{Reversing an \emph{even} quantity (Sec.~\ref{sec:TRS_breaking}), such as for instance $\ve{E}_0$, would have yielded relations of the kind $\tilde{\dya{\epsilon}}(\ve{E}_0)\neq\tilde{\dya{\epsilon}}^T(-\ve{E}_0)$, involving bias flipping and hence forbidding simultaneous excitations from both ends, since $\ve{E}_0$ cannot be \emph{simultaneously} pointing in opposite directions.\label{fn:OC_even}}. For instance, consider Eq.~\eqref{eq:Onsager_Casimir_rel_mu}. This time-reversal symmetric relation corresponds to \figref{fig:Gyro_Rec_Med}(b) with $\ve{B}_0=\ve{F}_0$, where the wave propagating in the direct direction (left picture) sees the medium $\tilde{\dya{\mu}}(\ve{F}_0)$, and the wave propagating in the reverse direction with flipped bias (right picture) sees the same effective medium $\tilde{\dya{\mu}}^T(-\ve{F}_0)=\tilde{\dya{\mu}}(\ve{F}_0)$ (although the \emph{system} has been altered). The negation of Eq.~\eqref{eq:Onsager_Casimir_rel_mu}, $\tilde{\dya{\mu}}(\ve{F}_0)\neq\tilde{\dya{\mu}}^T(+\ve{F}_0)$, is the time-reversal asymmetric relation corresponding to \figref{fig:Gyro_Rec_Med}(c), where the wave propagating in the direct direction (left picture) still sees the medium $\tilde{\dya{\mu}}(\ve{F}_0)$, whereas the wave propagating in the reverse direction with \emph{unflipped} bias (right picture) sees a \emph{different} effective medium $\tilde{\dya{\mu}}^T(+\ve{F}_0)\neq\tilde{\dya{\mu}}(\ve{F}_0)$ (without system alteration), consistently with the nonreciprocity of the system.

\section{Generality of Onsager-Casimir Relations}\label{sec:Gen_Onsager_Casimir_rel}

Although the frequency-domain derivation in Sec.~\ref{sec:gen_rec_th} assumed linear time-invariance, \textbf{the Onsager-Casimir relations are totally general}. Indeed, Onsager derived them for linear process without any other assumption than \emph{microscopic reversibility} and basic theorems from the general theory of fluctuations~\cite{Onsager_1931_I}, and these relation were later extended to nonlinear systems~\cite{Trzeciecki_2000}. Therefore, the Onsager-Casimir relations \textbf{hold not only for LTI systems}, as shown in Sec.~\ref{sec:gen_rec_th}, \textbf{but also for LTV and nonlinear systems}.

Systems that are \emph{not} media, but rather inhomogeneous structures, components or devices, cannot, of course, be characterized by medium parameters, but the Onsager-Casimir relations hold then in terms of \emph{extended S-parameters and S-matrix}, as will be shown in Sec.~\ref{sec:scat_pat_mod}, and particularly Eq.~\eqref{eq:Onsager_Sp}.

\section{Finer Classification of Nonreciprocal Systems}\label{sec:further_classif}

Section~\ref{sec:NR_def_class} established a first gross classification of nonreciprocity in terms of \emph{linear and nonlinear nonreciprocal systems}. However, Sec.~\ref{sec:Lin_NR_media} has revealed the necessity to \textbf{further divide the linear category into \emph{linear time-invariant (LTI) and linear time-variant (LTV) nonreciprocal systems}}, although Eq.~\eqref{eq:TR_bianisotropic_rel} and Secs.~\ref{sec:ser_rec_from_TRS}-\ref{sec:gen_rec_th} apply only to the former type\footnote{Note that LTV systems are often considered as nonlinear because they generate new frequencies as nonlinear systems. However, the distingo between the two types is important in the context of nonreciprocal systems, because the former supports the \emph{principle of superposition} whereas the latter does not, which leads to very distinct properties. Consider for instance a simple isotropic dielectric medium, characterized by the constitutive relation $\ve{D}=\epsilon\ve{E}$. Such a system is clearly \emph{nonlinear} if $\epsilon=\epsilon(\ve{E})$, since it excludes superposition from the fact that $\epsilon=\epsilon(\ve{E}_1+\ve{E}_2)$~\cite{Boyd_2008}. However, a \emph{linear}-TV system, namely a space-time modulated system, characterized by a constitutive relation of the type $\epsilon=\epsilon(t)\neq\epsilon(\ve{E})$ clearly supports superposition, although it produces new frequency, as will be seen later.}. Table~\ref{tab:classification} reflects this fact while summarizing the applicability of the main concepts studied in the paper to the resulting three different categories. In addition to mentioning the most common bias field involved, it presents \textbf{\emph{time reversal symmetry breaking} and \emph{Onsager-Casimir relations} in terms of generic S-matrix/tensor ($\dya{S}$) relations as a common descriptor for \emph{all} nonreciprocal systems}, stresses that the Lorentz theorem is applicable only to LTI systems, and indicates that extended S-parameters apply to the three types of nonreciprocal systems, with increasing restriction from LTI through LTV to nonlinear.

\begin{table}[h]
\footnotesize
\centering
\begin{tabular}{c|ccc}
& \multicolumn{2}{c}{\textbf{LINEAR}} & \textbf{NONLINEAR} \\
& \textsc{TI} & \textsc{TV (Space-Time)} & \\
 & \footnotesize{$\dya{S}(\ve{F}_0)\neq\dya{S}^T(\ve{F}_0)$}\:\phantom{.}
 & \footnotesize{$\dya{S}(\ve{F}_0)\neq\dya{S}^T(\ve{F}_0)$}
 & \footnotesize{$\dya{S}(\ve{F}_0)\neq\dya{S}^T(\ve{F}_0)$}\\
\\[-0.9em]
\hline
\\[-0.8em]
\textsc{Common bias $\ve{F}_0$} & magn. field ($\ve{B}_0$) & velocity ($\ve{v}_0$) & wave field ($\ve{E}$) \\
\\[0.1em]
\begin{minipage}{0.135\textwidth}\centering
\textsc{Time Reversal} \\
Secs.~\ref{sec:TRS}-\ref{sec:TRSB_example}
\end{minipage}
& \ding{52}& \ding{52} & \ding{52} \\
\\
\begin{minipage}{0.135\textwidth}\centering
\textsc{Onsager-Casimir} \\
Eqs.~\eqref{eq:Onsager_Casimir_rel}
\end{minipage}
& \ding{52} & \ding{52}  & \ding{52} \\
\\
\begin{minipage}{0.135\textwidth}\centering
\textsc{Lorentz NR} \\
Secs.~\ref{sec:Lin_NR_media}-\ref{sec:gen_rec_th}
\end{minipage}
& \ding{52} & \ding{56}  & \ding{56} \\
\\
\begin{minipage}{0.135\textwidth}\centering
\textsc{Extended S-Par.} \\
Sec.~\ref{sec:scat_pat_mod}
\end{minipage}
& \ding{52} & \ding{52}\danger  & \ding{52}\danger\danger
\end{tabular}
\caption{Fundamental concept applicability to the three different types of nonreciprocal (NR) systems: linear time invariant (TI), linear time variant (TV) and nonlinear.}
\label{tab:applicability}
\end{table}

\section{Reciprocity despite Time-Reversal Symmetry Breaking in Lossy Systems}\label{sec:Lossy_syst}

We now consider the case of lossy nonreciprocal systems, and particularly how to handle time reversal and time-reversal symmetry breaking in such systems, which is not trivial and possibly confusing.

\begin{figure}[h]
    \centering
        \psfragfig*[width=\linewidth]{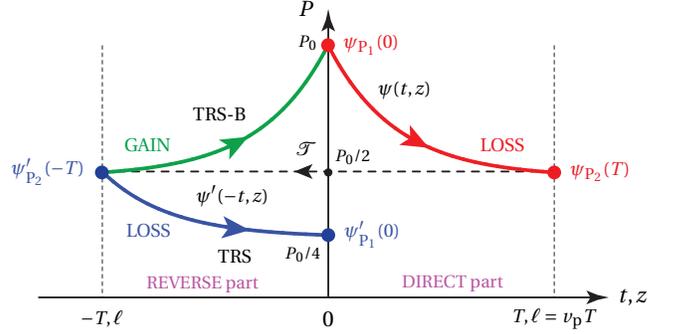}{
        \psfrag{t}[c][c]{$t,z$}
        \psfrag{P}[c][c]{$P$}
        \psfrag{0}[c][c]{$0$}
        \psfrag{T}[c][c][0.85]{$T,\ell=v_\tx{p}T$}
        \psfrag{M}[c][c][0.85]{$-T,\ell$}
        \psfrag{Q}[c][c][0.85]{$\psi(t,z)$}
        \psfrag{S}[c][c][0.85]{$\psi'(-t,z)$}
        \psfrag{Z}[c][c][0.75]{$P_0$}
        \psfrag{Y}[c][c][0.75]{$P_0/2$}
        \psfrag{X}[c][c][0.75]{$P_0/4$}
        \psfrag{N}[c][c][0.85]{\textcolor{red}{LOSS}}
        \psfrag{L}[c][c][0.85]{\textcolor{blue}{LOSS}}
        \psfrag{G}[c][c][0.85]{\textcolor{green}{GAIN}}
        \psfrag{1}[c][c][0.85]{\textcolor{red}{$\psi_{\tx{P}_1}(0)$}}
        \psfrag{2}[c][c][0.85]{\textcolor{red}{$\psi_{\tx{P}_2}(T)$}}
        \psfrag{3}[c][c][0.85]{\textcolor{blue}{$\psi_{\tx{P}_2}'(-T)$}}
        \psfrag{4}[c][c][0.85]{\textcolor{blue}{$\psi_{\tx{P}_1}'(0)$}}
        \psfrag{B}[c][c][0.85]{TRS}
        \psfrag{C}[c][c][0.85]{TRS-B}
        \psfrag{D}[c][c][0.8]{\textcolor{magenta}{DIRECT part}}
        \psfrag{E}[c][c][0.8]{\textcolor{magenta}{REVERSE part}}
        \psfrag{R}[c][c][0.85]{${\cal T}$}}
        \caption{Time-reversal symmetry breaking (TRS-B) in a lossy reciprocal waveguide of length $\ell$ (Sec.~\ref{sec:TRS_breaking}). Assuming that the process under consideration is the propagation of a modulated pulse, we have $\ves{\Psi}'(-t)=[\psi_\tx{P$_1$}'(-t),\psi_\tx{P$_2$}'(-t)]^T
        \neq[\psi_\tx{P$_1$}(t),\psi_\tx{P$_2$}(t)]^T=\ves{\Psi}(-t)$, and in particular, with the power loss assumed in the figure, $\ves{\Psi}'(0)=[P_0/4,0]^T\neq[P_0,0]^T=\ves{\Psi}(0)$.}
   \label{fig:Loss_TRSB}
\end{figure}

Figure~\ref{fig:Loss_TRSB} monitors the process of electromagnetic wave propagation in a simple \textbf{\emph{lossy bias-less waveguide system}}. Let us see how this system responds to time reversal (Secs.~\ref{sec:TRS}--\ref{sec:TRS_EM}) by applying the \textbf{time reversal symmetry breaking test} in Sec.~\ref{sec:TRS_breaking}, as for the Faraday system in Figs.~\ref{fig:Gyro_Rec_Med}(b),(c).

In the direct part of the process, the wave is attenuated by dissipation as it propagates from port P$_1$ to port P$_2$ (red curve), say from $P_0$ to $P_0/2$ (3~dB loss). Upon time reversal, the propagation direction is reversed, \emph{and} loss is transformed into gain (Tab.~\ref{tab:Field_Symmetries}). As a result, the wave propagates back from P$_2$ to P$_1$ \emph{and} its power level is restored (green curve), from $P_0/2$ to $P_0$. However, the \emph{system has been altered}. Maintaining it lossy leads to further attenuation on the return trip, from $P_0/2$ to $P_0/4$ (6~dB loss), and hence \textbf{breaks time-reversal symmetry}. According to Sec.~\ref{sec:TRS_breaking}, this would imply nonreciprocity, which is at odds with the generalized reciprocity theorem (Sec.~\ref{sec:gen_rec_th})!

This paradox originates from the looseness\footnote{This looseness has been tolerated because it is the \emph{ratio} definition of Sec.~\ref{sec:NR_def_class}, and not its restricted \emph{level} form, that is commonly used in practice.} of the assumption that \textbf{``Time-reversal symmetry/asymmetry is equivalent to reciprocity/nonreciprocity''} in Sec.~\ref{sec:TRS}. This assumption is, as pointed out in Fn.~\ref{fn:loose_def}, \emph{stricto senso} incorrect, as the equivalence \textbf{only holds in terms of absolute field levels} but not field ratios. In the case of loss, as just seen, the field/power ratios are equal [$(P_0/4)/(P_0/2)=0.5=(P_0/2)/P_0$], consistently with the general definition of reciprocity in Sec.~\ref{sec:NR_def_class}, but the field/power levels are not ($P_0/4\neq P_0$), in contradiction with the definition of time reversal in Sec.~\ref{sec:TRS}. In this sense, \textbf{\emph{a simple lossy system is perfectly reciprocal despite breaking time-reversal symmetry}}. This time-reversal asymmetry may be seen as an expression of \textbf{\emph{thermodynamical macroscopic irreversibility}}\footnote{Consider for instance an empty metallic waveguide. The transfer of charges from the waveguide lossy walls results in electromagnetic energy being transformed into heat (Joule first law)~\cite{Casimir_1963}. In theory, a \emph{Maxwell demon}~\cite{Knott_1911}
could reverse the velocities of all the molecules of the system, which would surely reconvert that heat into electromagnetic energy. In this sense, \emph{all systems are microscopically reversible}, which is the fundamental assumption underpinning Onsager reciprocity relations~\cite{Onsager_1931_I,Onsager_1931_II,Casimir_1945,Landau_1997} (Fn.~\ref{fn:Onsager}). However, such reconversion is prohibited by the second law of thermodynamics, which stipulates that \emph{the total entropy in an isolated system cannot decrease over time}. It would at the least require injecting energy from the outside of the system! So, such a lossy system is \emph{macroscopically -- and hence practically! -- irreversible}. Loss cannot be undone; it ever accumulates over time, as in \figref{fig:Loss_TRSB}.}.

A direct corollary of these considerations is that, while time-reversal symmetry necessarily implies reciprocity, time-reversal symmetry breaking does not necessarily imply nonreciprocity.

\section{Open Systems and their Time-Reversal Symmetry ``Lossy'' Behavior}\label{sec:open_syst}

\begin{figure}[h]
\centering
        \psfragfig*[width=\linewidth]{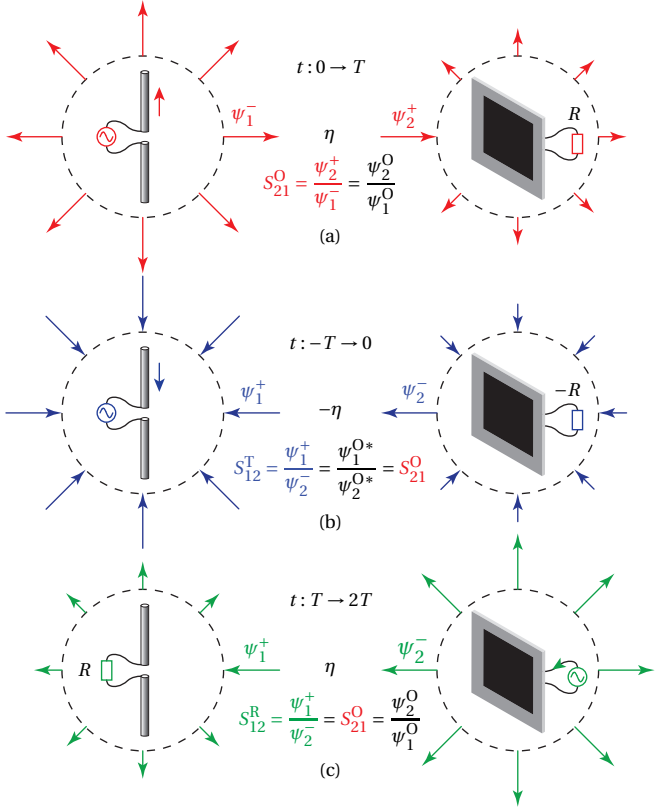}{
        \psfrag{a}[c][c][0.85]{(a)}
        \psfrag{b}[c][c][0.85]{(b)}
        \psfrag{c}[c][c][0.85]{(c)}
        \psfrag{1}[c][c][0.85]{$\eta$}
        \psfrag{2}[c][c][0.85]{$-\eta$}
        \psfrag{3}[c][c][0.85]{$\eta$}
        \psfrag{4}[c][c][0.85]{$R$}
        \psfrag{5}[c][c][0.85]{$-R$}
        \psfrag{6}[c][c][0.85]{$R$}
        \psfrag{h}[c][c][0.85]{$\textcolor{red}{\psi_1^-}$}
        \psfrag{i}[c][c][0.85]{$\textcolor{red}{\psi_2^+}$}
        \psfrag{j}[c][c][0.85]{$\textcolor{blue}{\psi_1^+}$}
        \psfrag{k}[c][c][0.85]{$\textcolor{blue}{\psi_2^-}$}
        \psfrag{l}[c][c][0.85]{$\textcolor{green}{\psi_1^+}$}
        \psfrag{m}[c][c]{$\textcolor{green}{\psi_2^-}$}
        \psfrag{e}[c][c][0.85]{$\textcolor{red}{S_{21}^\text{O}=\dfrac{\psi_2^+}{\psi_1^-}}=\dfrac{\psi_2^\text{O}}{\psi_1^\text{O}}$}
        \psfrag{f}[c][c][0.85]{$\textcolor{blue}{S_{12}^\text{T}=\dfrac{\psi_1^+}{\psi_2^-}}=\dfrac{\psi_1^{\text{O}*}}{\psi_2^{\text{O}*}}=\textcolor{red}{S_{21}^\text{O}}$}
        \psfrag{g}[c][c][0.85]{$\textcolor{green}{S_{12}^\text{R}=\dfrac{\psi_1^+}{\psi_2^-}}=\textcolor{red}{S_{21}^\text{O}}=\dfrac{\psi_2^\text{O}}{\psi_1^\text{O}}$}
        \psfrag{O}[c][c][0.85]{$t:0\rightarrow T$}
        \psfrag{T}[c][c][0.85]{$t:-T\rightarrow 0$}
        \psfrag{R}[c][c][0.85]{$t:T\rightarrow 2T$}
        }
        \caption{Time-reversal asymmetry and apparent (restricted) nonreciprocity of an open system composed of a dipole antenna and a patch antenna in free space.}
\label{fig:Diff_TRS_Rec_Open}
\end{figure}

Consider the two-antenna open system in \figref{fig:Diff_TRS_Rec_Open}, showing the original [Fig.~\ref{fig:Diff_TRS_Rec_Open}(a)], time-reversed [Fig.~\ref{fig:Diff_TRS_Rec_Open}(b)] and reciprocal [Fig.~\ref{fig:Diff_TRS_Rec_Open}(c)] problems. The nature of the system is clearly altered upon time reversal, where the intrinsic impedance of the surrounding medium becomes negative (Tab.~\ref{tab:Field_Symmetries}). This results from the fact that \textbf{the \emph{radiated and scattered energy escaping the antennas} in the original problem is \emph{equivalent to loss} relatively to the \emph{two-port system}}. Such loss transforms into gain upon time reversal, as in Sec.~\ref{sec:Lossy_syst}, leading to fields emerging from infinity. Upon replacing time reversal by restricted time reversal (Sec.~\ref{sec:ser_rec_from_TRS}) so as to avoid denaturing the system, one would find, as in the lossy case, reduced field levels but conserved field ratios\footnote{However, the restricted time-reversed problem is still \emph{distinct} from the reciprocity problem [Fig.~\ref{fig:Diff_TRS_Rec_Open}(c)] insofar as it does not reset the field level.}. \textbf{An open system is thus time-reversal-wise similar to a lossy system} (Sec.~\ref{sec:Lossy_syst}).

\section{Extended Scattering Parameter Modeling}\label{sec:scat_pat_mod}

The lossy/open system difficulty (Secs.~\ref{sec:Lossy_syst}/\ref{sec:open_syst}), the common definition in Sec.~\ref{sec:NR_def_class} and the general reciprocity relations derived by Onsager~\cite{Lewis_1925,Onsager_1931_I,Onsager_1931_II} suggest describing \emph{non/reciprocal systems} in terms of \textbf{\emph{field ratios}}. This leads to the \textbf{\emph{scattering parameters} or \emph{S-parameters}}, introduced in quantum physics in 1937~\cite{Wheeler_1937}, used for over 70 years in microwave engineering for LTI systems~\cite{Dicke_1947,Marcuvitz_1951,Pozar_ME_2011}, extended to power parameters for arbitrary loads in the 1960ies~\cite{Kurokawa_1965,Pozar_ME_2011}, and to the cross-coupled matrix theory for topologically-coupled resonators in the 2000s~\cite{Atia_1972,Cameron_2007}. We shall attempt here an \textbf{\emph{extension} of these parameters to LTV and nonlinear systems}.

\begin{figure}[h]
    \centering
        \psfragfig*[width=0.9\linewidth]{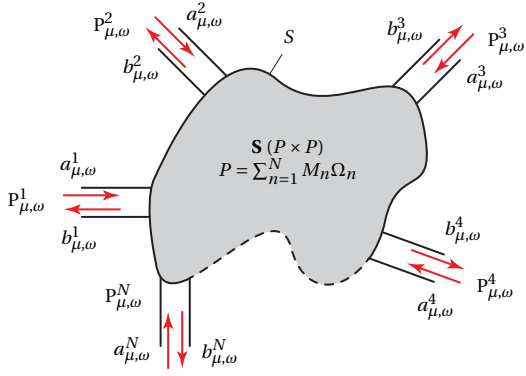}{
        \psfrag{S}[c][c][0.85]{\begin{minipage}{3cm}\centering $\ve{S}$ ($P\times P$) \\ \small$P=\sum_{n=1}^N M_n\Omega_n$ \end{minipage}}
        \psfrag{a}[c][c][0.85]{$a^1_{\mu,\omega}$}
        \psfrag{b}[c][c][0.85]{$b^1_{\mu,\omega}$}
        \psfrag{c}[c][c][0.85]{$a^2_{\mu,\omega}$}
        \psfrag{d}[c][c][0.85]{$b^2_{\mu,\omega}$}
        \psfrag{e}[c][c][0.85]{$a^3_{\mu,\omega}$}
        \psfrag{f}[c][c][0.85]{$b^3_{\mu,\omega}$}
        \psfrag{g}[c][c][0.85]{$a^4_{\mu,\omega}$}
        \psfrag{h}[c][c][0.85]{$b^4_{\mu,\omega}$}
        \psfrag{i}[c][c][0.85]{$a^N_{\mu,\omega}$}
        \psfrag{j}[c][c][0.85]{$b^N_{\mu,\omega}$}
        \psfrag{O}[c][c][0.85]{$S$}
        \psfrag{1}[c][c][0.85]{P$^1_{\mu,\omega}$}
        \psfrag{2}[c][c][0.85]{P$^2_{\mu,\omega}$}
        \psfrag{3}[c][c][0.85]{P$^3_{\mu,\omega}$}
        \psfrag{4}[c][c][0.85]{P$^4_{\mu,\omega}$}
        \psfrag{N}[c][c][0.85]{P$^N_{\mu,\omega}$}
        }
        \caption{Arbitrary $P$-port network and \emph{extended} scattering matrix, $\ve{S}$.}
        \label{fig:Multiport}
\end{figure}

Figure~\ref{fig:Multiport} defines an \emph{extended arbitrary $P$-port network} as an electromagnetic structure delimited by a surface $S$ with $N$ \emph{waveguide terminals}, $T_n$, supporting each a number of mode-frequency\footnote{``Frequency'' here refers to the new frequency set (possibly infinite or continuous) in LTV and nonlinear systems, practically restricted to discrete spectra.\label{fn:Sp_restr}} ports, P$_p=$P$^n_{\mu,\omega}$ with $p=1,2,\ldots,P$ (e.g. if $T_1$ is a waveguide with the $M_1=2$ modes TE$_{10}$ and TM$_{11}$ and the $\Omega_1=2$ frequencies $\omega$ and $2\omega$, it includes the $M_1\Omega_1=4$ ports P$_1=$P$^1_{\text{TE}_{10},\omega}$, P$_2=$P$^1_{\text{TE}_{10},2\omega}$, P$_3=$P$^1_{\text{TM}_{11},\omega}$ and P$_4=$P$^1_{\text{TM}_{11},2\omega}$.) The transverse fields in the waveguides have the \emph{frequency-domain form}~\cite{Marcuvitz_1951}, extended here to LTV and nonlinear systems,
\begin{equation}\label{eq:WG_Modes}
\begin{Bmatrix}
\tilde{\ve{E}}_{t,p}(x,y,z) \\
\tilde{\ve{H}}_{t,p}(x,y,z)
\end{Bmatrix}
=
\left(a_p e^{-j\beta_p z}\pm b_p e^{+j\beta_n z}\right)
\begin{Bmatrix}
\tilde{\ve{e}}_{t,p}(x,y) \\
\tilde{\ve{h}}_{t,p}(x,y)
\end{Bmatrix},
\end{equation}
where $\oiint_S(\tilde{\ve{e}}_{t,p}\times\tilde{\ve{h}}_{t,q})\cdot\uve{n}ds=2\delta_{pq}$\footnote{This relation also applies when $p$ and $q$ differ only by frequency (same terminal/mode) assuming narrow-band, and hence independent, port detectors.}, and where $a_p$/$b_p$ ($p=1,\dots,P$) are the port input/output complex wave amplitudes, related by the \textbf{\emph{extended S-matrix}}, $\ve{S}$, as
\begin{equation}\label{eq:b_eq_Sa}
\ve{b}=\ve{S}\ve{a},
\quad\text{with}\quad
\begin{Bmatrix}
\ve{b}=[b_1,b_2,\ldots,b_P]^T \\
\ve{a}=[a_1,a_2,\ldots,a_P]^T
\end{Bmatrix}.
\end{equation}

If the system is \textbf{\emph{linear}}, and hence \emph{superposition} applies, each entry of the matrix can be expressed by the simple transfer function $S_{ij}=b_i/a_j|_{a_k=0,k\neq j}$, which corresponds to the \textbf{conventional definition} of the $S$-parameters, except for the \textbf{frequency port definition extension in the space-time (LTV) case}. If the system is \textbf{\emph{nonlinear}}, then $S_{ij}=S_{ij}(a_1,a_2,\dots,a_P)$, and therefore \textbf{\emph{all} the (significant) input signals must be simultaneously present in the measurement} of the transfer function $S_{ij}$, as done in \emph{Broadband Poly-Harmonic Distortion (PHD)}, used in the Keysight microwave Nonlinear Vector Network Analyzer (NVNA)~\cite{Root_2005,Root_2008,PNAX_10_2017}\footnote{For instance, in a simple waveguide junction (without any bias), $S_{21}^\tx{junction}=b_2/a_1|_{a_2=a_3=0}\neq S_{21}(a_3)$, which allows measuring the system with separate inputs. In contrast, in a mixer (or modulator), $S_{21}^\tx{mixer}=S_{21}^\tx{mixer}(a_3=\tx{LO})$, and the local oscillator (or pump) port (P$_3$) must be excited along the the signal port (P$_1$) to determine the transfer function $S_{21}$.\label{fn:PNAX}}.

In an \emph{LTI medium}, the bianisotropic reciprocity relations~\cite{Kong_2008,Rothwell_2008}  excited by the two states $\ve{a}'$ and $\ve{a}''$ with responses $\ve{b}'$ and $\ve{b}''$ read now\footnote{The sources are now outside of the system and do therefore not contribute, contrary to the situation of Eq.~\eqref{eq:reac_EDHB}.}~\cite{Supp_Mat_NR}
\begin{gather}
\nabla\cdot\left(\tilde{\ve{E}}'\times\tilde{\ve{H}}''
-\tilde{\ve{E}}''\times\tilde{\ve{H}}'\right)
=j\omega\left[
\tilde{\ve{E}}''\cdot\left(\tilde{\dya{\epsilon}}
-\tilde{\dya{\epsilon}}^T\right)\cdot\tilde{\ve{E}}'\right.
-\tilde{\ve{H}}''\cdot\left(\tilde{\dya{\mu}}\right.
\nonumber \\
\left.-\tilde{\dya{\mu}}^T\right)\cdot\tilde{\ve{H}}'
+\tilde{\ve{E}}''\cdot\left(\tilde{\dya{\xi}}
+\tilde{\dya{\zeta}}^T\right)\cdot\tilde{\ve{H}}'\left.-\tilde{\ve{H}}''\cdot\left(\tilde{\dya{\zeta}}
+\tilde{\dya{\xi}}^T\right)\cdot\tilde{\ve{E}}'\right]=0.
\label{eq:gen_rec_the_for_S}
\end{gather}
Inserting the sum ($\sum_{p=1}^P$) of fields~\eqref{eq:WG_Modes} into this equation, taking the volume integral of the resulting relation, applying the Gauss theorem and the orthogonality relation, and using the Onsager-Casimir relations [Eqs.~\eqref{eq:Onsager_Casimir_rel}],
yields~\cite{Supp_Mat_NR}
\begin{equation}
\textstyle{\sum_p}(b_p'a_p''-a_p'b_p'')
=\ve{b}\ve{a}^{\prime\prime T}-\ve{a}\ve{b}^{\prime\prime T}
=\ve{a}'\ve{a}^{\prime\prime T}\left(\ve{S}^T-\ve{S}\right),
\end{equation}
where~\eqref{eq:b_eq_Sa} has been used to eliminate $\ve{b}^{(\prime,\prime\prime)}$ in the last equality. This leads to the reciprocity condition $\ve{S}=\ve{S}^T$, and hence to the convenient \textbf{scattering-parameter nonreciprocity condition}
\begin{equation}\label{eq:Sp_NR}
  \ve{S}\neq\ve{S}^T
  \;\text{or}\;\; \exists~(i,j)~|~
  S_{ij}\neq S_{ji},\,i=,1,2,\ldots,N,
\end{equation}
(e.g. $S_{21}\neq S_{12}$) which also applies to LTV and nonlinear systems, although the current demonstration is restricted to LTI media.


In the microwave regime, these S-parameters can be directly measured (magnitude and phase) with a VNA~\cite{Pozar_ME_2011} or with an NVNA. In contrast, in the optical regime no specific instrumentation is available for that and a special setup, with nontrivial phase handling, is therefore required~\cite{Jalas_2013}.

The symmetry reciprocity relation $\ve{S}=\ve{S}^T$ and nonreciprocity relation $\ve{S}\neq\ve{S}^T$ [Eq.~\eqref{eq:Sp_NR}], assuming the extended S-parameters introduced in this section, are nothing but the \textbf{general Onsager-Casimir reciprocity/nonreciprocity relations}~\cite{Onsager_1931_I,Onsager_1931_II,Casimir_1945,Casimir_1963}, and may be explicitly written as
\begin{subequations}\label{eq:Onsager_Sp}
\begin{equation}
\dya{S}(\ve{F}_0)=\dya{S}^T(-\ve{F}_0)\quad\tx{(reciprocity)},
\end{equation}
\begin{equation}
\dya{S}(\ve{F}_0)\neq\dya{S}^T(\ve{F}_0)\quad\tx{(nonreciprocity)},
\end{equation}
\end{subequations}
where the matrix $\ve{S}$ has been written in the tensor notation $\dya{S}$ for direct comparison with~\eqref{eq:Onsager_Casimir_rel}.

\section{Energy Conservation}\label{sec:energy_cons}

\textbf{In a \emph{lossless system}, \emph{energy conservation} requires that the total output power equals the total input power}, or $\sum_{p=1}^P|b_p|^2=\sum_{p=1}^P|a_p|^2$ in \figref{fig:Multiport}, since no power is dissipated in the system. In terms of \textbf{S-matrix}, this requirement translates into the \textbf{unitary} relation
\begin{equation}\label{eq:S_recipr}
\ve{S}\ve{S}^\dagger=I,
\end{equation}
where the $\dagger$ symbol represents the transpose conjugate and $I$ the unit matrix. Many fundamental useful facts on multi-port systems straightforwardly follow from energy conservation (e.g.~\cite{Pozar_ME_2011}).

Some immediate \textbf{consequences of Eq.~\eqref{eq:S_recipr} for nonreciprocity} are:
\begin{enumerate}
  \item \emph{A lossless \textbf{1-port system}, $\ve{S}=[S_{11}]$, can be only totally reflective}, from $|S_{11}|^2=1$, even if it includes nonreciprocal materials, contrary to claims in~\cite{Tsakmakidis_2017}.
  \item \emph{A \textbf{2-port system}, $\ve{S}=[S_{11},S_{12};S_{21},S_{22}]$, can be magnitude-wise nonreciprocal only if it is lossy}; specifically, a purely reflective 2-port isolator $\ve{S}=[0,0;1,1]$ is impossible, since energy conservation requires $|b_2|^2=|a_1|^2+|a_2|^2$, whereas the device would exhibit $b_2^2=a_1^2+a_2^2+2a_1a_2$, with the additional term $2a_1a_2$ that may cause the total energy to be larger than the input power, depending on the relative phases of $a_1$ and $a_2$.
  \item \emph{A lossless \textbf{2-port system} can still be nonreciprocal in terms of phase}, since Eq.~\eqref{eq:S_recipr} does \emph{not} demand $\angle S_{21}=\angle S_{12}$~\cite{Liberal_2014,Zhang_TMTT_09_2015}.
  \item \emph{A lossless \textbf{3-port system}, $\ve{S}=[S_{11},S_{12},S_{13};S_{21},S_{22},S_{23};$ $S_{31},S_{32},S_{33}]$, can be matched simultaneously at all ports only if it is nonreciprocal}, as may be shown by manipulating the matrix system~\eqref{eq:S_recipr}~\cite{Pozar_ME_2011}.
\end{enumerate}

\section{The ``Thermodynamics Paradox''}\label{sec:thermo_par}

The case 2) in Sec.~\ref{sec:energy_cons} -- the 2-port isolator -- has raised much perplexity in the past, and led to the so-called \textbf{\emph{``thermodynamics paradox.''}} The paradox states that an isolator system would ever increase the temperature of the load at the passing end at the detriment of the load at the isolated end, hence violating the second law of thermodynamics, which prescribes heat transfer from hot to cold bodies.

The paradox started in 1885 with the comment by Rayleigh that the recently developed system composed of two Nicols sandwiching a magnetized dielectric would be ``inconsistent with the second law of thermodynamics''~\cite{Strutt_1885}.

It was overruled 16 years later by Rayleigh himself~\cite{Rayleigh_1901}, who realized then, following related studies of Wiener~\cite{Wiener_1900}, a misunderstanding of the system, that was actually \emph{not} nonreciprocal, as the wave on the presumed stop-direction could eventually exit the device after three round-trips across the device (see Fn.~\ref{fn:no_isolator}).

The paradox resurfaced in 1955, as Lax and Button pointed out the existence of lossless unidirectional eigenmodes in some ferrite-loaded waveguide structures~\cite{Lax_1955}. It was eventually resolved by Ishimaru, who showed that such a waveguide would necessarily support substantial loss in its terminations, due to energy conservation (Sec.~\ref{sec:energy_cons}), even in the limit of negligible material loss~\cite{Ishimaru_1962,Ishimaru_1990}. This loss would eventually heat up the isolator, and therefore ensure thermal balance through thermal emission towards the cold bath. The paradox was due to the resolution of Maxwell equations for a completely lossless medium, which corresponds to an ``improperly posed problem,'' not corresponding to physical reality\footnote{Solving Maxwell equations for a medium with non zero conductivity and letting the conductivity go to zero completely resolves the issue~\cite{Ishimaru_1962,Ishimaru_1990}.}.

\section{Overview of Three Fundamental Types \\ of Nonreciprocal Systems}\label{sec:overview_three_NR_types}

At this point, the fundamental concepts of reciprocity have been covered to the level judged appropriate for such a review. Upon this foundation, we shall now describe the aforementioned three types of nonreciprocal systems (Tab.~\ref{tab:applicability}), namely LTI, LTV and nonlinear nonreciprocal systems, which will be covered in Secs.~\ref{sec:lin_NR_comp}, \ref{sec:ST_var_syst} and~\ref{sec:nonlin_syst}, respectively. In each case, we will list the fundamental characteristics, enumerate the main applications and describe a particular example.

\section{Linear-TI Nonreciprocal Systems}\label{sec:lin_NR_comp}

\emph{LTI nonreciprocal systems} have the following fundamental characteristics:
\begin{enumerate}
  \item Time-reversal symmetry breaking by time-reversal odd \textbf{external bias} $\ve{F}_0$, which is most often a \textbf{magnetic field, $\ve{B}_0$}, as for instance in Fig.~\ref{fig:Gyro_Rec_Med}(c);
  \item applicability, by linearity, to \textbf{arbitrary excitations and intensities, i.e. \emph{strong nonreciprocal}}, as announced in Tab.~\ref{tab:classification};
  \item \textbf{frequency conservation}, again by linearity, and hence \emph{unrestricted} frequency-domain description (Secs.~\ref{sec:Lin_NR_media} and \ref{sec:ser_rec_from_TRS}), and full applicability of the \textbf{Lorentz reciprocity theorem} (Sec.~\ref{sec:gen_rec_th}) and of the \textbf{S-parameter machinery} (Sec.~\ref{sec:scat_pat_mod});
  \item generally \textbf{based on LTI materials}~\cite{Lax_1962,Auracher_1975,Ishak_1988,Rodrigue_1988,Zvezdin_Kotov_1997,Adam_2002,Kodera_2009,Kodera_2010,Parsa_TAP_03_2011}, including 2DEGs and graphene~\cite{Sounas_APL_01_2011,Sounas_TMTT_04_2012,Chamanara_MWCL_07_2012,Chamanara_OE_05_2013,Sounas_APL_05_2013,Tamagnone_2014,Tamagnone_2016}, \textbf{or metamaterials}~\cite{Carignan_2009,Boucher_2009,Kodera_APL_07_2011,Carignan_2011,Kodera_AWPL_01_2012,Kodera_AWPL_12_2012,Wang_PNAS_2012,Sounas_TAP_01_2013,Kodera_TMTT_03_2013,Taravati_TAP_07_2017,Kodera_2018}\footnote{In the particular case of magnetic structures such as hexaferrites~\cite{Mahmood_Hexaferrites_2016} or ferromagnetic nano-particles membranes~\cite{Carignan_2009}, the external bias is applied \emph{before} operation (pre-magnetization) and maintained from the shape of the nano-particles constituting the magnetic material. So, these structures may be considered in a sense as ``magnetless.'' Unfortunately, there is a fundamental tradeoff between remanence, and hence isolation strength, and loss, and such materials are therefore not competitive with magnetic materials subjected to an external bias in operation.} (Sec.~\ref{sec:Lin_NR_media}).
\end{enumerate}

The main LTI nonreciprocal systems are isolators, nonreciprocal phase shifters and circulators~\cite{Lax_1962,Rodrigue_1988,Pozar_ME_2011}. \textbf{\emph{Isolators}} ($\ve{S}=[0,0;1,0]$) may be of \emph{Faraday, resonance, field-displacement or matched-port-circulator} type, and may involve resistive sheets, quarter-wave plates or polarizing grids. They are typically used to shield equipment (e.g. VNA or laser) from detuning, interfering and even destructive reflections. \textbf{\emph{nonreciprocal phase shifters}} ($\ve{S}=[0,e^{j\Delta\varphi};1,0]$), with the \emph{gyrator} ($\Delta\varphi=\pi$)~\cite{Tellegen_1948} as a particular case, may be of \emph{latching (hard magnetic hysteresis) or Faraday rotation} type, and may involve quarter-wave plates. They combine with couplers to form isolators or circulators, provide compact simulated inductors and filter inverters, and enable nonreciprocal pattern and scanning arrays. \textbf{\emph{Circulators}} ($\ve{S}=[0,0,1;1,0,0;0,1,0]$) may be of \emph{4-port Faraday rotation or 3-port junction rotation} type. They are used for isolation, duplexing (radar and communication), and reflection amplifiers.

Figure~\ref{fig:Gyro_Rec_Med} represents a Faraday rotator, whose operation has been described in Sec.~\ref{sec:TRSB_example}. This system constitutes the key building brick of a Faraday isolator. A typical implementation of such an isolator involves a Faraday rotation angle of $\phi=\pi/4$ and two linear polarizers sandwiching the magnetic (generally ferrite) medium, rotated $\phi=\pi/4$ with respect to each other. In the passing direction, the electric field at P$_1$ is perpendicular to the first polarizer grid and therefore fully traverses it; it is next rotated in the -- say clockwise -- direction by $\pi/4$, so as to emerge perpendicularly to the second grid, which leads to full transmission to P$_2$, corresponding to $S_{21}=1$. In the stopping direction, the wave enters the system at P$_2$, so as to perpendicularly face the second grid and hence completely traverse it; it is then rotated, still clockwise, by $\pi/4$. As a result, it is now parallel to the first grid, and hence fully reflected by it, which leads in principle\footnote{This ``stopping-direction'' trip of the wave may be tricky. Indeed, without proper precaution, the wave reflected by the first grid eventually reaches P$_1$ after an additional round trip, hence ruining the intended isolator operation of the device. This is in fact the point that was misunderstood by Rayleigh in 1885 with the functionally similar magnetized dielectric -- Nicols device, that was in reality reciprocal and hence representing no thermodynamics paradox (Sec.~\ref{sec:thermo_par}.) Proper precaution includes the addition of highly resistive sheets in cascade with the system, as shown in~\cite{Parsa_TAP_03_2011}.\label{fn:no_isolator}} to $S_{12}\approx 0$.

\section{Linear Time-Variant Space-Time Nonreciprocal Systems}\label{sec:ST_var_syst}

\emph{LTV space-time}\footnote{As will be seen later, LTV nonreciprocity requires also spatial inversion symmetry breaking.} \emph{nonreciprocal systems} have the following fundamental characteristics:
\begin{enumerate}
  \item Time-reversal symmetry breaking by time-reversal odd \textbf{external bias} ($\ve{F}_0$) \textbf{velocity, $\ve{v}_0$}, as will be seen in the example below;
  \item \textbf{strong nonreciprocity}, for the same reason as LTI systems (Sec.~\ref{sec:lin_NR_comp});
  \item \textbf{generation of new, possibly anharmonic\footnote{Since the modulation energy is external to the system, it can support any frequency, and hence lead to any frequency generation.}, frequencies}, and hence \emph{restricted} applicability of \textbf{S-parameters} (Fn.~\ref{fn:Sp_restr});
  \item \textbf{\emph{moving medium}} (moving matter, e.g. opto-mechanical)~\cite{Sommerfeld_1952,Pauli_1958,Kong_2008,Jackson_1998,Kippenberg_2007} \textbf{or \emph{moving wave}} (moving perturbation, e.g. electro/acousto/nonlinear-optic)\footnote{Moving and modulated media both produce Doppler shifts~\cite{Eden_Doppler_1992} and nonreciprocity. In contrast, only the former supports Fizeau drag~\cite{Jackson_1998,Fizeau_1851} and bianisotropy transformation~\cite{Kong_2008}, and only the latter allows superluminality~\cite{Deck_PRB_03_2018}.}~\cite{Cassedy_1963,Saleh_Teich_FP_2007};
  \item pulse or periodic and abrupt or smooth material/perturbation modulations.
\end{enumerate}

\begin{figure}[h]
\centering
\psfragfig*[width=\linewidth]{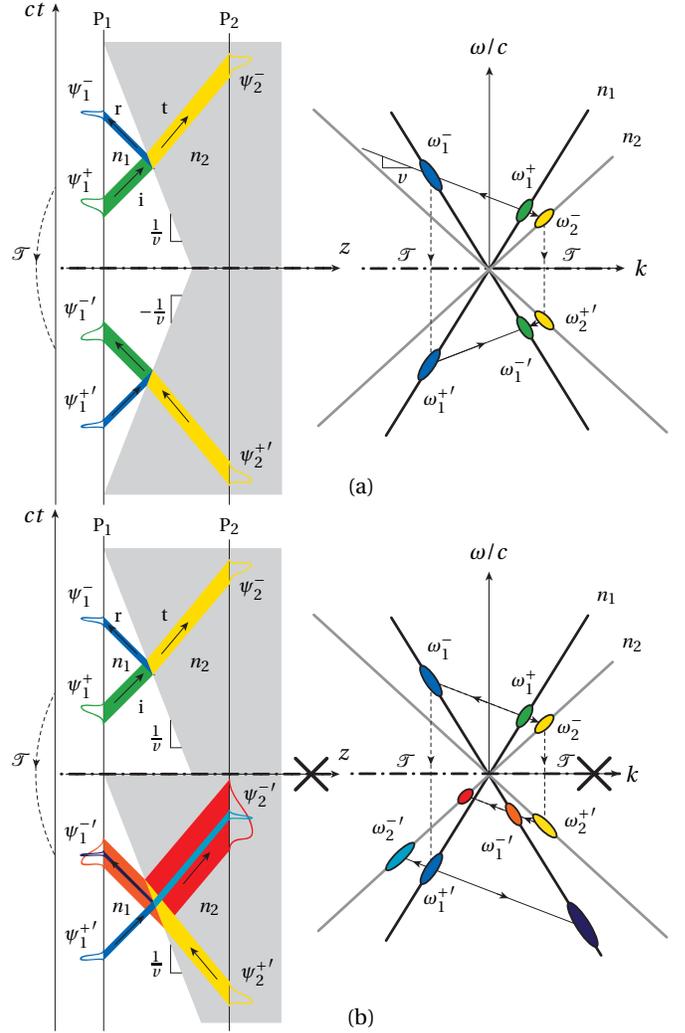}{
\psfrag{1}[c][c][0.85]{$n_1$}
\psfrag{2}[c][c][0.85]{$n_2$}
\psfrag{L}[c][c][0.85]{P$_1$}
\psfrag{R}[c][c][0.85]{P$_2$}
\psfrag{z}[c][c]{$z$}
\psfrag{Q}[c][c]{$ct$}
\psfrag{Y}[c][c]{(a)}
\psfrag{Z}[c][c]{(b)}
\psfrag{M}[c][c][0.85]{$n_1$}
\psfrag{N}[c][c][0.85]{$n_2$}
\psfrag{w}[c][c]{$\omega/c$}
\psfrag{k}[c][c]{$k$}
\psfrag{V}[c][c][0.85]{$\frac{1}{v}$}
\psfrag{v}[c][c][0.85]{$-\frac{1}{v}$}
\psfrag{s}[c][c][0.85]{$v$}
\psfrag{A}[c][c][0.85]{$\psi_1^+$}
\psfrag{B}[c][c][0.85]{$\psi_1^-$}
\psfrag{C}[c][c][0.85]{$\psi_2^-$}
\psfrag{a}[c][c][0.85]{${\psi_1^-}'$}
\psfrag{b}[c][c][0.85]{${\psi_1^+}'$}
\psfrag{c}[c][c][0.85]{${\psi_2^+}'$}
\psfrag{d}[c][c][0.85]{${\psi_2^-}'$}
\psfrag{E}[c][c][0.85]{$\omega_1^+$}
\psfrag{F}[c][c][0.85]{$\omega_1^-$}
\psfrag{G}[c][c][0.85]{$\omega_2^-$}
\psfrag{e}[c][c][0.85]{${\omega_1^-}'$}
\psfrag{f}[c][c][0.85]{${\omega_1^+}'$}
\psfrag{g}[c][c][0.85]{${\omega_2^+}'$}
\psfrag{h}[c][c][0.85]{${\omega_2^-}'$}
\psfrag{T}[c][c][0.8]{${\cal T}$}
\psfrag{i}[c][c][0.85]{i}
\psfrag{r}[c][c][0.85]{r}
\psfrag{t}[c][c][0.85]{t}
        }
\caption{Step space-time modulated system, $t>0$. (a)~$t<0$: time-reversal symmetric. (b)~$t<0$: time-reversal asymmetric -- nonreciprocal.}
\label{fig:Minkowski_TR}
\end{figure}

As an illustration, Fig.~\ref{fig:Minkowski_TR} graphically depicts, using an extended Minkowski diagram representation~\cite{Deck_PRB_03_2018}, a \textbf{step space-time modulated system} with interface between media of refractive indices $n_1$ and $n_2$ ($n_2>n_1$) moving in the $-z$-direction with the constant velocity $\ve{v}_0=v\uve{z}$ ($v<0$) and excited by an incident (i) wave propagating in the $+z$-direction. Upon full time reversal, $\ve{v}_0$ is reversed, which leads to identical \emph{Doppler shifts}~\cite{Eden_Doppler_1992} in the reflected (r) and transmitted (t) waves\footnote{Temporal frequency ($\omega$) and spatial frequency ($\ve{k}$) transitions follow corresponding frequency conservation lines in the moving frame~\cite{Deck_PRB_03_2018}.}, as in Fig.~\ref{fig:Minkowski_TR}(a), but alters the system. The unaltered system is time-reversal asymmetric, and hence breaks time-reversal symmetry, which leads to the nonreciprocal scattering in Fig.~\ref{fig:Minkowski_TR}(b)\footnote{Purely temporal modulation (horizontal interface between the two (white and gray) media)~\cite{Kalluri_ETVCM_2010,Fang_2012,Fang_2013} would clearly be insufficient for nonreciprocity; spatial inversion symmetry breaking, provided here by the moving modulation, is also required to break time-reversal symmetry: $\dya{\chi}(\ve{v}_0)\neq\dya{\chi}^T(\ve{v}_0)$.}.

A great diversity of useful space-time modulated nonreciprocal systems have been reported in recent years~\cite{Winn_1999,Dong_2008,Yu_2009,Yu_2009_2,Manipatruni_2009,Kang_2011,Yu_2011,Kamal_2011,Doerr_2011,Sounas_NATCOM_09_2013,Galland_2013,Qin_TMTT_2014,Sounas_ACS_2014,Estep_2014,Hadad_2015,Hadad_2016,Reiskarimian_2016,Taravati_TAP_02_2017,Taravati_PRB_03_2018,Li_04_2018,Attarzadeh_2018}. They are all based on the production of \textbf{\emph{different traveling phase gradient in opposite directions}} and are, in that sense, more or less lumped/distributed~\cite{Pozar_ME_2011} \emph{variations of parametric systems developed by microwave engineers in the 1950ies}\footnote{The main difference is that the space-time modulated or \emph{parametric} systems of that time were developed mostly for amplifiers or mixers, rather than nonreciprocal devices.}~\cite{Pierce_1950,Cullen_1958,Tien_1958,Tien_1958_2,Landauer_1960,LePage_1953,Franks_1960}, but they may lead to many novel structures and applications, especially when more than one spatial dimension is involved.

\begin{figure}[h]
    \centering
        \psfragfig*[width=\linewidth]{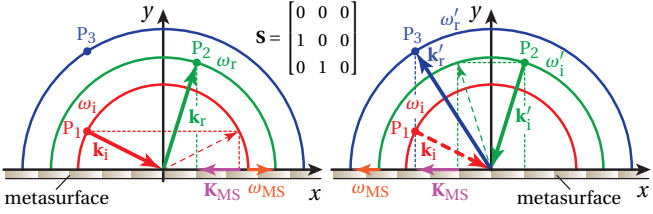}{
        \psfrag{x}[c][c]{$x$}
        \psfrag{y}[c][c]{$y$}
        \psfrag{1}[c][c][0.85]{\color{red}{$\ve{k}_\text{i}$}}
        \psfrag{2}[c][c][0.85]{\color{green}{$\ve{k}_\text{r}$}}
        \psfrag{3}[c][c][0.85]{\color{green}{$\ve{k}_\text{i}'$}}
        \psfrag{4}[c][c][0.85]{\color{blue}{$\ve{k}_\text{r}'$}}
        \psfrag{5}[c][c][0.85]{\color{magenta}{$\ve{K}_\text{MS}$}}
        \psfrag{6}[c][c][0.85]{\color{red}{$\omega_\text{i}$}}
        \psfrag{7}[c][c][0.85]{\color{green}{$\omega_\text{r}$}}
        \psfrag{8}[c][c][0.85]{\color{green}{$\omega_\text{i}'$}}
        \psfrag{9}[c][c][0.85]{\color{blue}{$\omega_\text{r}'$}}
        \psfrag{d}[c][c][0.85]{\color{orange}{$\omega_\text{MS}$}}
        \psfrag{h}[c][c][0.85]{\color{orange}{$\omega_\text{MS}$}}
        \psfrag{a}[c][c]{}
        \psfrag{b}[c][c]{}
        \psfrag{k}[c][c][0.85]{\color{red}{P$_1$}}
        \psfrag{l}[c][c][0.85]{\color{green}{P$_2$}}
        \psfrag{m}[c][c][0.85]{\color{blue}{P$_3$}}
        \psfrag{M}[c][c][0.85]{metasurface}
        \psfrag{S}[c][c][0.8]{$\ve{S}=\begin{bmatrix}0&0&0\\1&0&0\\0&1&0\end{bmatrix}$}
        }
        \caption{Three-port nonreciprocal space-time modulated reflective metasurface with extended $\ve{S}$ matrix.}
        \label{fig:ST_Metasurface}
\end{figure}

Figure~\ref{fig:ST_Metasurface} shows such a multi-dimensional system, specifically a \textbf{\emph{nonreciprocal metasurface reflector}}~\cite{Shaltout_OME_11_2015,Hadad_2015} based on the space-time modulation $n(x)=n_0+n_\tx{m}\cos(\beta_\tx{m}x+\omega_\tx{m}t)$, where $(\omega_\tx{m}/\beta_\tx{m})\uve{z}=\ve{v}_0$. The space-time modulated metasurface breaks reciprocity and hence provides a quite unique nonreciprocal device by adding the spatial and temporal momenta $\ve{K}_\tx{MS}$ and $\omega_\tx{MS}$ to those of the incident wave\footnote{This system actually includes infinitely many ports, $S_{n,n+1}=0$ and $S_{n+1,n}$ with $\omega_{n+1}>\omega_n$ ($n=1,\ldots\infty$), but its \emph{functional} reduction in the figure is meaningful if the power transfer beyond P$_3$ is of no interest.}.

\section{Nonlinear Nonreciprocal Systems}\label{sec:nonlin_syst}

\emph{Nonlinear nonreciprocal} systems have the following fundamental characteristics:
\begin{enumerate}
  \item Time-reversal symmetric breaking by \textbf{\emph{spatial asymmetry} and \emph{nonlinear self-biasing}} (nonlinearity triggering by wave itself)~\cite{Naguleswaran_1998,Trzeciecki_2000}, as will be seen in the forthcoming example;
  \item limitation to \textbf{restricted excitations, intensities and isolation, i.e. \emph{weak nonreciprocity}}, as will also be understood from that example;
  \item \textbf{generation of new, only harmonic\footnote{The restriction to harmonic frequencies is a consequence of the assumed \emph{self-biasing} nature of nonlinear nonreciprocal systems (Fn.~\ref{fn:no_ext_bias_NL}).}, frequencies, and inapplicability of superposition}, and hence \emph{very restricted} applicability of \textbf{S-parameters} (Fns.~\ref{fn:Sp_restr} and~\ref{fn:PNAX});
  \item large diversity of possible time-reversal symmetry breaking approaches.
\end{enumerate}

\begin{figure}[h]
    \centering
        \psfragfig*[width=\linewidth]{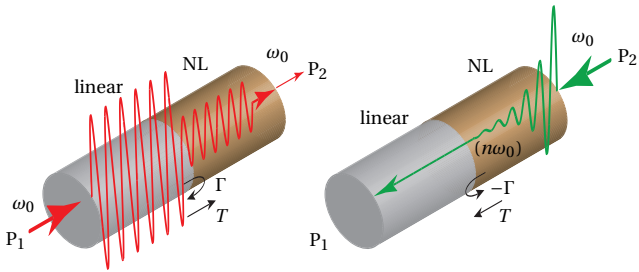}{
        \psfrag{N}[l][l][0.85]{NL}
        \psfrag{L}[l][l][0.85]{linear}
        \psfrag{X}[c][c][0.85]{P$_1$}
        \psfrag{Y}[c][c][0.85]{P$_2$}
        \psfrag{w}[c][c][0.85]{$\omega_0$}
        \psfrag{z}[c][c][0.85]{$(n\omega_0)$}
        \psfrag{R}[c][c][0.85]{$\Gamma$}
        \psfrag{r}[c][c][0.85]{$-\Gamma$}
        \psfrag{T}[c][c][0.85]{$T$}
        }
        \caption{Principle of nonlinear (NL) nonreciprocity with the asymmetric cascade of a linear medium, $\epsilon_\tx{L}=\epsilon_1$, and a nonlinear lossy medium, $\epsilon_\tx{NL}=\epsilon_2+[\epsilon'(E)-j\epsilon''(E)]$, with high mismatch, $|\Gamma|=|(\sqrt{\epsilon_\tx{NL}}-\sqrt{\epsilon_\tx{L}})/(\sqrt{\epsilon_\tx{NL}}+\sqrt{\epsilon_\tx{L}})|\gg 0$.}
   \label{fig:NL_NR}
\end{figure}

Figure~\ref{fig:NL_NR} shows a simple way to achieve nonlinear nonreciprocity \textbf{by pairing a linear medium and a nonlinear lossy medium}. The two media are strongly \emph{mismatched}, with reflection~$\Gamma$. A wave injected at port P$_1$  experiences a transmittance of $|T|^2=1-|\Gamma|^2\ll 1$, yielding a much smaller field level in the nonlinear medium. If this level is insufficient to trigger nonlinear loss, all the power transmitted through the interface ($|T|^2$) reaches port P$_2$, so that $|S_{21}|=|T|$ and $|S_{11}|=|\Gamma|$. The same wave injected at P$_2$, assuming sufficient intensity to trigger nonlinear loss, undergoes exponential attenuation $e^{-\alpha\ell_\tx{NL}}$ ($\ell_\tx{NL}$: nonlinear length), so that $|S_{12}|=|T|e^{-\alpha\ell_\tx{NL}}\approx 0$. The system is thus nonreciprocal, but it is a \textbf{\emph{pseudo}-isolator}\footnote{The device is also \emph{not a diode}~\cite{Sze_Ng_PSD_2006}, whose nonreciprocity consists in different forward/backward spectra due to positive/negative wave cycle clipping.}: 1)~it is \textbf{restricted to a small range of intensities}; 2)~it works \textbf{only for one excitation direction} ($\tx{P}_1\rightarrow\tx{P}_2$ \emph{or} $\tx{P}_2\rightarrow\tx{P}_1$) \textbf{at a time}, since the $\tx{P}_2\rightarrow\tx{P}_1$ wave would trigger nonlinear loss and hence also extinct the $\tx{P}_1\rightarrow\tx{P}_2$ wave\footnote{This precludes most of the applications of real isolators (Sec.~\ref{sec:lin_NR_comp})~\cite{Shi_2015}.}; 3)~it often suffers from \textbf{poor isolation} ($|S_{21}|/|S_{12}|=e^{\alpha\ell_\tx{NL}}$) and \textbf{poor isolation to insertion loss ratio} ($(|S_{21}|/|S_{12}/|)/|S_{11}|=e^{\alpha\ell_\tx{NL}}/|\Gamma|$)\footnote{These two parameters are typically smaller than $20$dB/$25$dB in nonreciprocal nonlinear structures, whereas they commonly exceed $45$dB/$50$dB in nonreciprocal LTI isolators.}; 4)~it is \textbf{reciprocal to noise}~\cite{Shi_2015}.

Ingenious variations of the nonlinear nonreciprocal device in \figref{fig:NL_NR} have been reported~\cite{Gallo_1999,Kivshar_2010,Lepri_2011,Fan_2011,Fan_2012,Bender_2013,Wang_2013,Chang_2014,Nazari_2014,Engheta_2015,Bino_02_2017,Sounas_2018,Bino_2018,Sounas_AWPL_2018}. Some of them mitigate some of the aforementioned issues, but these improvements are severely restricted by fundamental limitations of nonlinear nonreciprocity~\cite{Jalas_2013,Shi_2015,Sounas_03_2018}.

\section{Distinction with Asymmetric Propagation}\label{sec:dist_asym}

\textbf{A nonreciprocal system is a system that exhibits time-reversal asymmetric field ratios between well-defined ports} [Eq.~\eqref{eq:Sp_NR}], which is possible only under \textbf{external biasing (linear nonreciprocity) \emph{or} self-biasing plus spatial asymmetry (nonlinear nonreciprocity)}. Any system not satisfying this condition is \emph{necessarily reciprocal}, despite possible \textbf{\emph{fallacious asymmetries in transmission}}\cite{Fang_1996,Wang_OE_2011,Feng_Science_2011,Fan_Science_2012,Wang_SR_2012,Jalas_2013,Maznev_2013,Shi_2015,Fernandez_AWPL_2018}.

\begin{figure}[h]
    \centering
        \psfragfig*[width=0.9\linewidth]{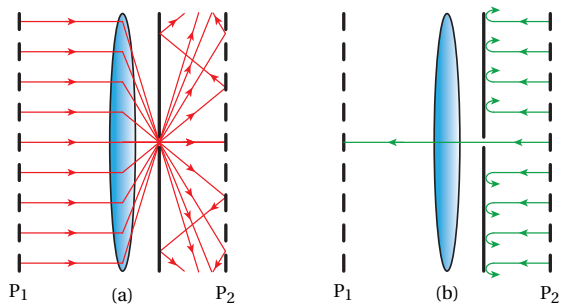}{
        \psfrag{a}[c][c][0.85]{(a)}
        \psfrag{b}[c][c][0.85]{(b)}
        \psfrag{x}[c][c][0.85]{P$_1$}
        \psfrag{y}[c][c][0.85]{P$_2$}
        }
        \caption{Asymmetric \emph{reciprocal} system formed by a lens and a mirror sandwiched between two antenna arrays under (a)~left and (b)~right excitations.}
   \label{fig:Lens_Hole_Asymmetry}
\end{figure}

For instance, the system in \figref{fig:Lens_Hole_Asymmetry} exhibits \textbf{asymmetric ray propagation}, but it is fully reciprocal since only the horizontal ray gets transmitted between the array ports, the $\tx{P}_1\rightarrow\tx{P}_2$ oblique rays symmetrically canceling out on the right array due to opposite phase gradients.

Other deceptively nonreciprocal cases include: \textbf{asymmetric field rotation-filtering} (e.g. $\pi/2$ rotator $+$ polarizer), where $S_{21}^{yx}\neq S_{21}^{xy}$, but $S_{21}^{yx}=S_{12}^{xy}$; \textbf{asymmetric waveguide junction} (e.g. step width variation), with full transmission to larger side distributed over multi-modes and small transmission from the same mode to the smaller (single-mode) side, but reciprocal mode-to-mode transmission (e.g. $S_{51}=S_{15}$, 1: small side single-mode port, 2: large side mode/port 5)~\cite{Jalas_2013}; \textbf{asymmetric mode conversion} (e.g. waveguide with nonuniform load), where even mode transmits in opposite directions with and without excitation of odd mode, without breaking reciprocity, since $S_{21}^\tx{ee}=S_{12}^\tx{ee}$ and $S_{21}^\tx{oe}=0=S_{12}^\tx{eo}$~\cite{Fan_Science_2012}. In all cases, reciprocity is verified upon exchanging the source and detector.

\section{Conclusion}\label{sec:concl}

This paper has presented, in the context of magnetless nonreciprocal systems aiming at repelling the frontiers of nonreciprocity technology, a global perspective of nonreciprocity, with the following main conclusions:
\begin{itemize}
  \item Nonreciprocal systems are defined as systems that exhibit different received-transmitted field ratios when their source(s) and detector(s) are exchanged.
  \item Nonreciprocity is equivalent to time-reversal symmetry breaking, except in the case where loss and/or gain are specifically considered, in which case nonrecipocity is expressed, despite broken time-reversal symmetry, in the restricted terms of equal fields ratios, consistently with the definition of the previous point.
  \item Time reversal / time-reversal symmetry breaking is a fundamental and powerful common descriptor for all nonreciprocal systems.
  \item In the case of LTI media, the time reversal leads to generalized Lorentz reciprocity theorem for nonreciprocal systems, leading to the Onsager-Casimir relations, from which nonreiprocity is expressed by the fact that waves propagating in opposite directions see different media for a fixed time-reversal odd bias.
  \item However, Onsager-Casimir relations are completely general, as shown by Onsager from microscopic irreversibility, applying hence also to LTV and nonlinear nonreciprocal systems; as our general definition, they are expressed in terms of field ratios, or transfer functions.
  \item Nonreciprocal systems may be classified into linear (TI and TV space-time modulated) and nonlinear systems, based on time-reversal symmetry breaking by external biasing and self-biasing plus spatial asymmetry, respectively.
  \item Nonlinear nonreciprocity is a weaker form of nonreciprocity than linear nonreciprocity, as it suffers from restricted intensities, one-way-at-a-time excitations, and often leads to poor isolation.
  \item S-parameters can be advantageously generalized to all types of nonreciprocal systems, with some important restrictions.
  \item Care must be exercised to avoid confusing asymmetric transmission with nonreciprocity in some fallacious systems.
\end{itemize}

Nonreciprocity is a rich and fascinating concept, that has and will continue to open new scientific and technological horizons.

\bibliography{PRAppl_NONRECIPROCITY_Caloz}

\clearpage
\section{Supplementary Material}

\subsection{TRS of Maxwell equations}

\noindent $\rightarrow$ Eq.~\cite{Supp_Mat_NR}.\eqref{eq:Maxwell}

\noindent $\bullet$ TR version of Maxwell equations:
\begin{subequations}\label{eq:Maxwell_TR_A}
\begin{equation}
\nabla\times\ve{E}^{\prime}=-\partial\ve{B}^{\prime}/\partial t^{\prime}
\end{equation}
\begin{equation}
\nabla\times\ve{H}^{\prime}=\partial\ve{D}^{\prime}/\partial t^{\prime}+\ve{J^{\prime}}
\end{equation}
\end{subequations}
\noindent $\bullet$ Applying the TR rules in Tab.~\cite{Supp_Mat_NR}.\ref{tab:Field_Symmetries} to~\eqref{eq:Maxwell_TR_A}:
\begin{subequations}\label{eq:Maxwell_TR_2}
\begin{equation}
\nabla\times(\ve{E})=-\partial(-\ve{B})/\partial(-t)
\end{equation}
\vspace{-5mm}
\begin{equation}
\nabla\times(-\ve{H})=\partial(\ve{D})/\partial (-t)+(-\ve{J})
\end{equation}
\end{subequations}
\noindent $\bullet$ Simplification of~\eqref{eq:Maxwell_TR_2}:
\begin{subequations}\label{eq:Maxwell_A}
\begin{equation}
\nabla\times\ve{E}=-\partial\ve{B}/\partial t
\end{equation}
\vspace{-5mm}
\begin{equation}
\nabla\times\ve{H}=\partial\ve{D}/\partial t+\ve{J}
\end{equation}
\end{subequations}
\noindent$\rightarrow$ identical to original Maxwell equation $\equiv$ proof of .~\cite{Supp_Mat_NR}.\eqref{eq:Maxwell}

\subsection{Frequency-Domain Expressions of TR Constitutive Parameters}

\noindent $\rightarrow$ Eq.~\cite{Supp_Mat_NR}.\eqref{eq:freq_const_rel}

\noindent $\bullet$ Frequency-domain constitutive relations for a biased LTI bianisotropic medium of~\cite{Supp_Mat_NR}.\eqref{eq:TR_bianisotropic_rel}:
\begin{subequations}\label{eq:TR_bianisotropic_rel_dir}
\begin{equation}
\tilde{\ve{D}}
=\tilde{\dya{\epsilon}}(\ve{F}_0)\cdot\tilde{\ve{E}}
+\tilde{\dya{\xi}}(\ve{F}_0)\cdot\tilde{\ve{H}}
\end{equation}
\begin{equation}
\tilde{\ve{B}}
=\tilde{\dya{\zeta}}(\ve{F}_0)\cdot\tilde{\ve{E}}
+\tilde{\dya{\mu}}(\ve{F}_0)\cdot\tilde{\ve{H}}
\end{equation}
\end{subequations}
\noindent $\bullet$ Same for the corresponding TR medium:
\begin{subequations}\label{eq:TR_bianisotropic_rel_TR}
\begin{equation}
\tilde{\ve{D}}^\prime
=\tilde{\dya{\epsilon}}^\prime(-\ve{F}_0)\cdot\tilde{\ve{E}}^\prime
+\tilde{\dya{\xi}}^\prime(-\ve{F}_0)\cdot\tilde{\ve{H}}^\prime
\end{equation}
\begin{equation}
\tilde{\ve{B}}^{(\prime)}
=\tilde{\dya{\zeta}}^\prime(-\ve{F}_0)\cdot\tilde{\ve{E}}^\prime
+\tilde{\dya{\mu}}^\prime(-\ve{F}_0)\cdot\tilde{\ve{H}}^\prime
\end{equation}
\end{subequations}
where the sign of $F_0$ has been reversed because this quantity is assumed to be TR-odd (Sec.~\cite{Supp_Mat_NR}.\ref{sec:Lin_NR_media}).

\noindent $\bullet$ Applying~\cite{Supp_Mat_NR}.\eqref{eq:TR_Pw} with TR rules in Tab.~\cite{Supp_Mat_NR}.\ref{tab:Field_Symmetries} to~\eqref{eq:TR_bianisotropic_rel_TR}:
\begin{subequations}\label{eq:TR_bianisotropic_rel_TR_1}
\begin{equation}
\tilde{\ve{D}}^*
=\tilde{\dya{\epsilon}}^\prime(\ve{F}_0)\cdot\tilde{\ve{E}}^*
+\tilde{\dya{\xi}}^\prime(\ve{F}_0)\cdot(-\tilde{\ve{H}}^*)
\end{equation}
\begin{equation}
(-\tilde{\ve{B}}^*)
=\tilde{\dya{\zeta}}^\prime(\ve{F}_0)\cdot\tilde{\ve{E}}^*
+\tilde{\dya{\mu}}^\prime(\ve{F}_0)\cdot(-\tilde{\ve{H}}^*)
\end{equation}
\end{subequations}
\noindent $\bullet$ Simplifying~\eqref{eq:TR_bianisotropic_rel_TR_1}:
\begin{subequations}\label{eq:TR_bianisotropic_rel_TR_2}
\begin{equation}
\tilde{\ve{D}}^*
=\tilde{\dya{\epsilon}}^\prime(\ve{F}_0)\cdot\tilde{\ve{E}}^*
-\tilde{\dya{\xi}}^\prime(\ve{F}_0)\cdot\tilde{\ve{H}}^*
\end{equation}
\begin{equation}
\tilde{\ve{B}}^*
=-\tilde{\dya{\zeta}}^\prime(\ve{F}_0)\cdot\tilde{\ve{E}}^*
+\tilde{\dya{\mu}}^\prime(\ve{F}_0)\cdot\tilde{\ve{H}}^*
\end{equation}
\end{subequations}
\noindent $\bullet$ Complex conjugate of~\eqref{eq:TR_bianisotropic_rel_TR_2}:
\begin{subequations}\label{eq:TR_bianisotropic_rel_TR_3}
\begin{equation}
\tilde{\ve{D}}
=\tilde{\dya{\epsilon}}^{\prime*}(-\ve{F}_0)\cdot\tilde{\ve{E}}
-\tilde{\dya{\xi}}^{\prime*}(-\ve{F}_0)\cdot\tilde{\ve{H}}
\end{equation}
\begin{equation}
\tilde{\ve{B}}
=-\tilde{\dya{\zeta}}^{\prime*}(-\ve{F}_0)\cdot\tilde{\ve{E}}
+\tilde{\dya{\mu}}^{\prime*}(-\ve{F}_0)\cdot\tilde{\ve{H}}
\end{equation}
\end{subequations}
\noindent $\bullet$ Comparing~\eqref{eq:TR_bianisotropic_rel_TR_3} and~\eqref{eq:TR_bianisotropic_rel_dir}:
\begin{subequations}\label{eq:TR_bianisotropic_rel_TR_4}
\begin{equation}
\tilde{\dya{\epsilon}}(\ve{F}_0)=\tilde{\dya{\epsilon}}^{\prime*}(-\ve{F}_0)\
\quad
\tilde{\dya{\xi}}(\ve{F}_0)=-\tilde{\dya{\xi}}^{\prime*}(-\ve{F}_0)\
\end{equation}
\begin{equation}
\tilde{\dya{\zeta}}(\ve{F}_0)=-\tilde{\dya{\zeta}}^{\prime*}(-\ve{F}_0)\
\quad
\tilde{\dya{\mu}}(\ve{F}_0)=\tilde{\dya{\mu}}^{\prime*}(-\ve{F}_0)
\end{equation}
\end{subequations}

\noindent $\bullet$ Complex conjugate and variable change $\ve{F}\rightarrow-\ve{F}_0$ in~\eqref{eq:TR_bianisotropic_rel_TR_4}:
\begin{subequations}\label{eq:TR_bianisotropic_rel_TR_5}
\begin{equation}
\tilde{\dya{\epsilon}}^\prime(\ve{F}_0)=\tilde{\dya{\epsilon}}^{*}(-\ve{F}_0)\
\quad
\tilde{\dya{\xi}}^\prime(\ve{F}_0)=-\tilde{\dya{\xi}}^{*}(-\ve{F}_0)\
\end{equation}
\begin{equation}
\tilde{\dya{\zeta}}^\prime(\ve{F}_0)=-\tilde{\dya{\zeta}}^{*}(-\ve{F}_0)\
\quad
\tilde{\dya{\mu}}^\prime(\ve{F}_0)=\tilde{\dya{\mu}}^{*}(-\ve{F}_0)
\end{equation}
\end{subequations}
\noindent $\equiv$ Eq.~\cite{Supp_Mat_NR}.\eqref{eq:freq_const_rel}

\noindent $\bullet$ Restricted-TR version of~\eqref{eq:TR_bianisotropic_rel_TR_5} (same without $^*$):

\begin{subequations}\label{eq:TR_bianisotropic_rel_reducedTR}
\begin{equation}
\tilde{\dya{\epsilon}}^\prime(\ve{F}_0)=\tilde{\dya{\epsilon}}(-\ve{F}_0)\
\quad
\tilde{\dya{\xi}}^\prime(\ve{F}_0)=-\tilde{\dya{\xi}}(-\ve{F}_0)\
\end{equation}
\begin{equation}
\tilde{\dya{\zeta}}^\prime(\ve{F}_0)=-\tilde{\dya{\zeta}}(-\ve{F}_0)\
\quad
\tilde{\dya{\mu}}^\prime(\ve{F}_0)=\tilde{\dya{\mu}}(-\ve{F}_0)
\end{equation}
\end{subequations}

\subsection{Frequency-Domain Expressions of TR Maxwell Equations}\label{sec:freq_TR_Max}

\noindent$\rightarrow$ Used in Eq.~\cite{Supp_Mat_NR}.\eqref{eq:reac_EDHB}

\noindent $\bullet$ Fourier transforming~\eqref{eq:Maxwell_A}:
\begin{subequations}\label{eq:Maxwell_c}
\begin{equation}
\nabla\times\tilde{\ve{E}}=-j\omega\tilde{\ve{B}}
\end{equation}
\begin{equation}
\nabla\times\tilde{\ve{H}}=j\omega\tilde{\ve{D}}+\tilde{\ve{J}}
\end{equation}
\end{subequations}

\noindent $\bullet$ Substituting $t'=-t$ in~\eqref{eq:Maxwell_TR_A}:
\begin{subequations}\label{eq:Maxwell_TR_mt}
\begin{equation}
\nabla\times\ve{E}^{\prime}=\partial\ve{B}^{\prime}/\partial t
\end{equation}
\begin{equation}
\nabla\times\ve{H}^{\prime}=-\partial\ve{D}^{\prime}/\partial t+\ve{J^{\prime}}
\end{equation}
\end{subequations}
\noindent $\bullet$ Fourier transforming~\eqref{eq:Maxwell_TR_mt}:
\begin{subequations}\label{eq:Maxwell_TR_FT}
\begin{equation}
\nabla\times\tilde{\ve{E}}^{\prime}=j\omega\tilde{\ve{B}}^{\prime}
\end{equation}
\begin{equation}
\nabla\times\tilde{\ve{H}}^{\prime}=-j\omega\tilde{\ve{D}}^{\prime}+\tilde{\ve{J}}^{\prime}
\end{equation}
\end{subequations}
\noindent $\bullet$ Applying~\cite{Supp_Mat_NR}.\eqref{eq:TR_Pw} with TR rules in Tab.~\cite{Supp_Mat_NR}.\ref{tab:Field_Symmetries} to~\eqref{eq:Maxwell_TR_FT}:
\begin{subequations}\label{eq:Maxwell_TR_FT_np_cc}
\begin{equation}
\nabla\times\tilde{\ve{E}}^*=-j\omega\tilde{\ve{B}}^*
\end{equation}
\begin{equation}
\nabla\times\tilde{\ve{H}}^*=j\omega\tilde{\ve{D}}^*+\tilde{\ve{J}}^*
\end{equation}
\end{subequations}
\noindent $\bullet$ Comparing~\eqref{eq:Maxwell_TR_FT_np_cc} and~\eqref{eq:Maxwell_c}: \emph{The frequency-domain Maxwell equations of the TR problem are identical to those of the original problem with all the field quantities conjugated. With this result, one simply obtains Eq.~\cite{Supp_Mat_NR}.\eqref{eq:reac_EDHB} following the usual reciprocity theorem procedure  using no stars for the first set of sources-responses and stars for the second one.}

\subsection{Derivation of}

\noindent$\rightarrow$ Eq.~\cite{Supp_Mat_NR}.\eqref{eq:reac_EDHB}

\noindent $\bullet$ Frequency-domain Maxwell equations for set of original (excitation-response) fields:
\begin{subequations}\label{eq:Maxwell_TR_FT_1}
\begin{equation}\label{eq:Maxwell_TR_FT_1a}
\nabla\times\tilde{\ve{E}}=-j\omega\tilde{\ve{B}}
\end{equation}
\begin{equation}
\nabla\times\tilde{\ve{H}}=j\omega\tilde{\ve{D}}+\tilde{\ve{J}}
\end{equation}
\end{subequations}

\noindent $\bullet$ Same for TR fields, using the final result of Sec.~\ref{sec:freq_TR_Max}:
\begin{subequations}\label{eq:Maxwell_FT_1}
\begin{equation}
\nabla\times\tilde{\ve{E}}^*=-j\omega\tilde{\ve{B}}^*
\end{equation}
\begin{equation}\label{eq:Maxwell_FT_1b}
\nabla\times\tilde{\ve{H}}^*=j\omega\tilde{\ve{D}}^*+\tilde{\ve{J}}^*
\end{equation}
\end{subequations}
\noindent $\bullet$ Subtracting~\eqref{eq:Maxwell_TR_FT_1a} dot multiplied by $\ve{H}^*$ from~\eqref{eq:Maxwell_FT_1b}
dot multiplied by $\ve{E}$, and doing the same with swapped non-superscripts and primes:
\begin{subequations}
\begin{equation}
\tilde{\ve{E}}\cdot\nabla\times\tilde{\ve{H}}^*-\tilde{\ve{H}}^*\cdot\nabla\times\tilde{\ve{E}}
=j\omega\tilde{\ve{E}}\cdot\tilde{\ve{D}}^*
+\tilde{\ve{E}}\cdot\tilde{\ve{J}}^*
+j\omega\tilde{\ve{H}}^*\cdot\tilde{\ve{B}}
\end{equation}
\begin{equation}
\tilde{\ve{E}}^*\cdot\nabla\times\tilde{\ve{H}}-\tilde{\ve{H}}\cdot\nabla\times\tilde{\ve{E}}^*
=j\omega\tilde{\ve{E}}^*\cdot\tilde{\ve{D}}
+\tilde{\ve{E}}^*\cdot\tilde{\ve{J}}
+j\omega\tilde{\ve{H}}\cdot\tilde{\ve{B}}^*
\end{equation}
\end{subequations}
and, applying the identity $\ve{A}\cdot\nabla\times\ve{B}-\ve{B}\cdot\nabla\times\ve{A}=-\nabla\cdot(\ve{A}\times\ve{B})$:
\begin{subequations}
\begin{equation}\label{eq:LmanipaF1}
-\nabla\cdot(\tilde{\ve{E}}\times\tilde{\ve{H}}^*)
=j\omega\tilde{\ve{E}}\cdot\tilde{\ve{D}}^*
+\tilde{\ve{E}}\cdot\tilde{\ve{J}}^*
+j\omega\tilde{\ve{H}}^*\cdot\tilde{\ve{B}}
\end{equation}
\begin{equation}\label{eq:LmanipbF1}
-\nabla\cdot(\tilde{\ve{E}}^*\times\tilde{\ve{H}})
=j\omega\tilde{\ve{E}}^*\cdot\tilde{\ve{D}}
+\tilde{\ve{E}}^*\cdot\tilde{\ve{J}}
+j\omega\tilde{\ve{H}}\cdot\tilde{\ve{B}}^*
\end{equation}
\end{subequations}

\noindent $\bullet$ Subtracting~\eqref{eq:LmanipaF1} from~\eqref{eq:LmanipbF1}:
\begin{equation}\label{eq:LcomplTD1}
\begin{split}
&\nabla\cdot(\tilde{\ve{E}}\times\tilde{\ve{H}}^*-\tilde{\ve{E}}^*\times\tilde{\ve{H}})\\
&\quad=j\omega\left(\tilde{\ve{E}}^*\cdot\tilde{\ve{D}}
-\tilde{\ve{E}}\cdot\tilde{\ve{D}}^*
+\tilde{\ve{H}}\cdot\tilde{\ve{B}}^*
-\tilde{\ve{H}}^*\cdot\tilde{\ve{B}}
\right)
\\
&\qquad+\tilde{\ve{E}}^*\cdot\tilde{\ve{J}}
-\tilde{\ve{E}}\cdot\tilde{\ve{J}}^*
\end{split}
\end{equation}
or
\begin{equation}\label{eq:LcomplTD1x}
\begin{split}
&\tilde{\ve{E}}^*\cdot\tilde{\ve{J}}
-\tilde{\ve{E}}\cdot\tilde{\ve{J}}^*
\\
&\quad=\nabla\cdot(\tilde{\ve{E}}\times\tilde{\ve{H}}^*-\tilde{\ve{E}}^*\times\tilde{\ve{H}}) \\
&\qquad-j\omega\left(\tilde{\ve{E}}^*\cdot\tilde{\ve{D}}
-\tilde{\ve{E}}\cdot\tilde{\ve{D}}^*
+\tilde{\ve{H}}\cdot\tilde{\ve{B}}^*
-\tilde{\ve{H}}^*\cdot\tilde{\ve{B}}
\right)
\end{split}
\end{equation}
\noindent $\bullet$ Integrating over the volume $V$ formed by the surface $S$ and applying the Gauss theorem:
\begin{equation}
\small
\begin{split}
\iiint_{V\rightarrow V_J}\tilde{\ve{J}}\cdot\tilde{\ve{E}}^*dv
-\iiint_{V\rightarrow V_{J}}\tilde{\ve{J}}^*\cdot\tilde{\ve{E}}dv
=\oiint_S\left(\tilde{\ve{E}}\times\tilde{\ve{H}}^*
-\tilde{\ve{E}}^*\times\tilde{\ve{H}}\right)\cdot\uve{n}ds \\
-j\omega\iiint_V\left(\tilde{\ve{E}}^*\cdot\tilde{\ve{D}}
-\tilde{\ve{E}}\cdot\tilde{\ve{D}}^*+\tilde{\ve{H}}\cdot\tilde{\ve{B}}^*
-\tilde{\ve{H}}^*\cdot\tilde{\ve{B}}\right)dv.
\end{split}
\end{equation}
$\equiv$ Eq.~\cite{Supp_Mat_NR}.\eqref{eq:reac_EDHB}

\subsection{Vanishing of the Surface Integral in}
\label{eq:IS0}

Eq.~\cite{Supp_Mat_NR}.\eqref{eq:reac_EDHB}

\noindent $\bullet$ The surface integral in Eq.~\cite{Supp_Mat_NR}.\eqref{eq:reac_EDHB} reads
\begin{equation}\label{eq:IS}
I_S=\oiint_S\left(\tilde{\ve{E}}\times\tilde{\ve{H}}^*
-\tilde{\ve{E}}^*\times\tilde{\ve{H}}\right)\cdot\uve{n}ds,
\end{equation}
\noindent $\bullet$ From the identity $(\ve{a}\times\ve{b})\cdot\ve{c}=\ve{c}\cdot(\ve{a}\times\ve{b})=\ve{a}\cdot(\ve{b}\times\ve{c})=\ve{b}\cdot(\ve{c}\times\ve{a})$:
\begin{subequations}\label{eq:integrand_IS}
\begin{equation}
(\tilde{\ve{E}}\times\tilde{\ve{H}}^*)\cdot\uve{n}
=\tilde{\ve{E}}\cdot(\tilde{\ve{H}}^*\times\uve{n})
=\tilde{\ve{H}}^*\cdot(\uve{n}\times\tilde{\ve{E}})
\end{equation}
\begin{equation}
(\tilde{\ve{E}^*}\times\tilde{\ve{H}})\cdot\uve{n}
=\tilde{\ve{E}^*}\cdot(\tilde{\ve{H}}\times\uve{n})
=\tilde{\ve{H}}\cdot(\uve{n}\times\tilde{\ve{E}}^*)
\end{equation}
\end{subequations}
\noindent $\bullet$ Impenetrable [Perfect Electric Conducteur (PEC) or Perfect Magnetic Conductor {PMC} or combination of the two] cavity:
\begin{subequations}\label{eq:imp_cav}
\begin{equation}
\left[\tilde{\ve{H}}\times\uve{n}\right]_{S=\tx{PMC}}
=\left[\tilde{\ve{H}}^*\times\uve{n}\right]_{S=\tx{PMC}}
=0
\end{equation}
or
\begin{equation}
\left[\uve{n}\times\tilde{\ve{E}}\right]_{S=\tx{PEC}}
=\left[\uve{n}\times\tilde{\ve{E}}^*\right]_{S=\tx{PEC}}
=0
\end{equation}
\end{subequations}

\noindent $\bullet$ Inserted~\eqref{eq:imp_cav} into~\eqref{eq:integrand_IS} and substituting the result into~\eqref{eq:IS}:
\begin{equation}
I_S=0
\end{equation}

\noindent $\bullet$ In unbounded medium, at an infinite distance from the source(s), the field is a plane wave:
\begin{equation}\label{eq:PW}
\left[\uve{n}\times\tilde{\ve{E}}=\eta\ve{H}\right]_{S=\infty}
\quad
\tx{and}
\quad
\left[\uve{n}\times\tilde{\ve{E}}^*=\eta\ve{H}^*\right]_{S=\infty}
\end{equation}
\noindent where the \emph{restricted-TR} assumed in Sec.~\cite{Supp_Mat_NR}.\ref{sec:gen_rec_th} has been used in the latter equation by \emph{not} changing $\eta$ into $\eta^*$

\noindent $\bullet$ Inserting~\eqref{eq:PW} into~\eqref{eq:integrand_IS} and substituting the result into the integrand of ~\eqref{eq:IS}:
\begin{equation}\label{eq:EHEsHH2eta}
\begin{split}
&\left[(\tilde{\ve{E}}\times\tilde{\ve{H}}^*
-\tilde{\ve{E}}^*\times\tilde{\ve{H}})\cdot\uve{n}\right]_{S=\infty} \\
&\qquad=\left[\tilde{\ve{H}}^*\cdot(\uve{n}\times\tilde{\ve{E}})
-\tilde{\ve{H}}\cdot(\uve{n}\times\tilde{\ve{E}}^*)
\right]_{S=\infty} \\
&\qquad\overset{\tx{by }\eqref{eq:PW}}{=}
\left[\tilde{\ve{H}}^*\cdot\eta\tilde{\ve{H}}
-\tilde{\ve{H}}\cdot\eta\tilde{\ve{H}}^*
\right]_{S=\infty} \\
&\qquad=|\ve{H}|^2(\eta-\eta)=0
\end{split}
\end{equation}

\noindent $\bullet$ Inserting~\eqref{eq:EHEsHH2eta}
inserted into~\eqref{eq:IS}:
\begin{equation}
I_S=0
\end{equation}

\subsection{Frequency-Domain Expressions of Constitutive Relations}\label{sec:Is0}

\noindent$\rightarrow$ Used in Eq.~\cite{Supp_Mat_NR}.\eqref{eq:gen_Lorentz}

\noindent $\bullet$ Frequency-domain constitutive relations: Eq.~\eqref{eq:TR_bianisotropic_rel_dir}
\begin{subequations}\label{eq:TR_bianisotropic_rel_dir_x}
\begin{equation}
\tilde{\ve{D}}
=\tilde{\dya{\epsilon}}(\ve{F}_0)\cdot\tilde{\ve{E}}
+\tilde{\dya{\xi}}(\ve{F}_0)\cdot\tilde{\ve{H}}
\end{equation}
\begin{equation}
\tilde{\ve{B}}
=\tilde{\dya{\zeta}}(\ve{F}_0)\cdot\tilde{\ve{E}}
+\tilde{\dya{\mu}}(\ve{F}_0)\cdot\tilde{\ve{H}}
\end{equation}
\end{subequations}
\noindent $\bullet$ Frequency-domain \emph{restricted-TR} constitutive relations -- Eq.~\eqref{eq:TR_bianisotropic_rel_reducedTR} into Eq.~\eqref{eq:TR_bianisotropic_rel_TR_2}:
\begin{subequations}\label{eq:TR_bianisotropic_rel_TR_x}
\begin{equation}
\tilde{\ve{D}}^*
=\tilde{\dya{\epsilon}}(-\ve{F}_0)\cdot\tilde{\ve{E}}^*
+\tilde{\dya{\xi}}(-\ve{F}_0)\cdot\tilde{\ve{H}}^*
\end{equation}
\begin{equation}
\tilde{\ve{B}}^*
=\tilde{\dya{\zeta}}(-\ve{F}_0)\cdot\tilde{\ve{E}}^*
+\tilde{\dya{\mu}}(-\ve{F}_0)\cdot\tilde{\ve{H}}^*
\end{equation}
\end{subequations}
\noindent $\bullet$ Comparing~\eqref{eq:TR_bianisotropic_rel_TR_x} and~\eqref{eq:TR_bianisotropic_rel_dir_x}: \emph{The frequency-domain constitutive relations of the TR problem are identical to those of the original problem with all the field (but not constitutive parameter!) quantities conjugated and $\ve{F}_0$ changed to $-\ve{F}_0$.}

\subsection{Derivation of the Generalized Lorentz Theorem}\label{sec:GLT_A}

\noindent$\rightarrow$ Eq.~\cite{Supp_Mat_NR}.\eqref{eq:gen_Lorentz}

\noindent $\bullet$ Using the result of the vanishing of surface integral in Sec.~\ref{eq:IS0} and the fact that the LHS (reaction difference) in~\cite{Supp_Mat_NR}.\eqref{eq:reac_EDHB} vanishes under reciprocity, Eq.~\cite{Supp_Mat_NR}.\eqref{eq:reac_EDHB} reduces to:
\begin{subequations}
\end{subequations}
\begin{equation}\label{eq:vol_int}
\iiint_V X\,dv=0,
\end{equation}
\text{with}
\begin{equation}\label{eq:X_vol_eq}
\ve{X}=\tilde{\ve{E}}^*\cdot\tilde{\ve{D}}
-\tilde{\ve{E}}\cdot\tilde{\ve{D}}^*+\tilde{\ve{H}}\cdot\tilde{\ve{B}}^*
-\tilde{\ve{H}}^*\cdot\tilde{\ve{B}}
\end{equation}
\begin{subequations}
\end{subequations}
\noindent $\bullet$ Substituting~\eqref{eq:TR_bianisotropic_rel_dir_x} and~\eqref{eq:TR_bianisotropic_rel_TR_x} into \eqref{eq:X_vol_eq}:
\begin{equation}\label{eq:X_DB_subs}
\begin{split}
\ve{X}
=&\tilde{\ve{E}}^*\cdot\left[\tilde{\dya{\epsilon}}(\ve{F}_0)\cdot\tilde{\ve{E}}
+\tilde{\dya{\xi}}(\ve{F}_0)\cdot\tilde{\ve{H}}\right]\\
&-\tilde{\ve{E}}\cdot\left[\tilde{\dya{\epsilon}}(-\ve{F}_0)\cdot\tilde{\ve{E}}^*
+\tilde{\dya{\xi}}(-\ve{F}_0)\cdot\tilde{\ve{H}}^*\right] \\
&+\tilde{\ve{H}}\cdot\left[\tilde{\dya{\zeta}}(-\ve{F}_0)\cdot\tilde{\ve{E}}^*
+\tilde{\dya{\mu}}(-\ve{F}_0)\cdot\tilde{\ve{H}}^*\right] \\
&-\tilde{\ve{H}}^*\cdot\left[\tilde{\dya{\zeta}}(\ve{F}_0)\cdot\tilde{\ve{E}}
+\tilde{\dya{\mu}}(\ve{F}_0)\cdot\tilde{\ve{H}}\right] \\
=&\tilde{\ve{E}}^*\cdot\tilde{\dya{\epsilon}}(\ve{F}_0)\cdot\tilde{\ve{E}}
-\tilde{\ve{E}}\cdot\tilde{\dya{\epsilon}}(-\ve{F}_0)\cdot\tilde{\ve{E}}^* \\
&+\tilde{\ve{H}}\cdot\tilde{\dya{\mu}}(-\ve{F}_0)\cdot\tilde{\ve{H}}^*
-\tilde{\ve{H}}^*\cdot\tilde{\dya{\mu}}(\ve{F}_0)\cdot\tilde{\ve{H}} \\
&+\tilde{\ve{E}}^*\cdot\tilde{\dya{\xi}}(\ve{F}_0)\cdot\tilde{\ve{H}}
+\tilde{\ve{H}}\cdot\tilde{\dya{\zeta}}(-\ve{F}_0)\cdot\tilde{\ve{E}}^* \\
&+\tilde{\ve{H}}\cdot\dya{\zeta}(-\ve{F}_0)\cdot\tilde{\ve{E}}^*
+\tilde{\ve{E}}^*\cdot\tilde{\dya{\xi}}(\ve{F}_0)\cdot\tilde{\ve{H}}
\end{split}
\end{equation}
\noindent $\bullet$ Applying the tensor identity $\ve{a}\cdot\dya{\chi}\cdot\ve{b}
=\left(\ve{a}\cdot\dya{\chi}\cdot\ve{b}
\right)^T=\ve{b}\cdot\dya{\chi}^T\cdot\ve{a}$ (scalar quantity) to the terms at the right in the last equality of~\eqref{eq:X_DB_subs}:
\begin{equation}\label{eq:X_DB_subs_id}
\begin{split}
\ve{X}
=&\tilde{\ve{E}}^*\cdot\left[\tilde{\dya{\epsilon}}(\ve{F}_0)-\tilde{\dya{\epsilon}}^T(-\ve{F}_0)\right]\cdot\tilde{\ve{E}} \\
&+\tilde{\ve{H}}\cdot\left[\tilde{\dya{\mu}}(-\ve{F}_0)-\tilde{\dya{\mu}}^T(\ve{F}_0)\right]\cdot\tilde{\ve{H}}^* \\
&+\tilde{\ve{E}}^*\cdot\left[\tilde{\dya{\xi}}(\ve{F}_0)+\tilde{\dya{\zeta}}^T(-\ve{F}_0)\right]\cdot\tilde{\ve{H}} \\
&+\tilde{\ve{H}}\cdot\left[\dya{\zeta}(-\ve{F}_0)+\tilde{\dya{\xi}}^T(\ve{F}_0)\right]\cdot\tilde{\ve{E}}^*
\end{split}
\end{equation}
\noindent $\bullet$ Inserting~\eqref{eq:X_DB_subs_id} into~\eqref{eq:vol_int}, and considering that the resulting relation must hold for any fields:
\begin{subequations}
\vspace{-1mm}
\begin{equation}
\tilde{\dya{\epsilon}}(\ve{F}_0)
=\tilde{\dya{\epsilon}}^T(-\ve{F}_0),
\end{equation}
\vspace{-5mm}
\begin{equation}
\tilde{\dya{\mu}}(\ve{F}_0)=\tilde{\dya{\mu}}^T(-\ve{F}_0),
\end{equation}
\vspace{-5mm}
\begin{equation}
\tilde{\dya{\xi}}(\ve{F}_0)
=-\tilde{\dya{\zeta}}^T(-\ve{F}_0).
\end{equation}
\end{subequations}
$\equiv$ Eq.~\cite{Supp_Mat_NR}.\eqref{eq:gen_Lorentz}.

\subsection{Derivation of the LTI Medium S-Matrix NR/Reciprocity Condition}

\noindent$\rightarrow$ Eq.~\cite{Supp_Mat_NR}.\eqref{eq:gen_rec_the_for_S}

\noindent $\bullet$ Eq.~\eqref{eq:LcomplTD1} with non-superscripts replaced by primes and stars replaced by double primes:
\begin{equation}\label{eq:nEpdp}
\begin{split}
&\nabla\cdot(\tilde{\ve{E}}'\times\tilde{\ve{H}}''-\tilde{\ve{E}}''\times\tilde{\ve{H}}')\\
&\quad=j\omega\left(\tilde{\ve{E}}''\cdot\tilde{\ve{D}}'
-\tilde{\ve{E}}'\cdot\tilde{\ve{D}}''
+\tilde{\ve{H}}'\cdot\tilde{\ve{B}}''
-\tilde{\ve{H}}''\cdot\tilde{\ve{B}'}
\right)
\\
&\qquad+\tilde{\ve{E}}''\cdot\tilde{\ve{J}}'
-\tilde{\ve{E}}'\cdot\tilde{\ve{J}}''
\end{split}
\end{equation}

\noindent $\bullet$ Considering that the domain of interest, i.e. integration, will not include the sources, as in Sec.~\cite{Supp_Mat_NR}.\ref{sec:scat_pat_mod}, and hence dropping the source terms in Eq.~\eqref{eq:nEpdp}:
\begin{equation}\label{eq:nEpdp2}
\begin{split}
&\nabla\cdot(\tilde{\ve{E}}'\times\tilde{\ve{H}}''-\tilde{\ve{E}}''\times\tilde{\ve{H}}')\\
&\quad=j\omega\left(\tilde{\ve{E}}''\cdot\tilde{\ve{D}}'
-\tilde{\ve{E}}'\cdot\tilde{\ve{D}}''
+\tilde{\ve{H}}'\cdot\tilde{\ve{B}}''
-\tilde{\ve{H}}''\cdot\tilde{\ve{B}'}
\right)
\end{split}
\end{equation}
\noindent $\bullet$ Applying in~\eqref{eq:nEpdp2} similar substitutions as in Sec.~\ref{sec:GLT_A}:
\begin{equation}\label{eq:gen_rec_the_for_S_App}
\begin{split}
\nabla\cdot&\left(\tilde{\ve{E}}'\times\tilde{\ve{H}}''
-\tilde{\ve{E}}''\times\tilde{\ve{H}}'\right) \\
=&j\omega\left[\tilde{\ve{E}}''\cdot\left(\tilde{\dya{\epsilon}}
-\tilde{\dya{\epsilon}}^T\right)\cdot\tilde{\ve{E}}'\right.
-\tilde{\ve{H}}''\cdot\left(\tilde{\dya{\mu}}-\tilde{\dya{\mu}}^T\right)\cdot\tilde{\ve{H}}' \\
&\left.+\tilde{\ve{E}}''\cdot\left(\tilde{\dya{\xi}}+\tilde{\dya{\zeta}}^T\right)\cdot\tilde{\ve{H}}'
-\tilde{\ve{H}}''\cdot\left(\tilde{\dya{\zeta}}
+\tilde{\dya{\xi}}^T\right)\cdot\tilde{\ve{E}}'\right]=0.
\end{split}
\end{equation}
$\equiv$ Eq.~\cite{Supp_Mat_NR}.\eqref{eq:gen_rec_the_for_S}

\subsection{Derivation of the S-Matrix NR/Reciprocity Condition}

\noindent$\rightarrow$ Eq.~\cite{Supp_Mat_NR}.\eqref{eq:Sp_NR}

\noindent $\bullet$ In~\eqref{eq:gen_rec_the_for_S_App} [Eq.~\cite{Supp_Mat_NR}.\eqref{eq:gen_rec_the_for_S}] elimination of RHS from general Onsager-Casimir relations [Eq.~\cite{Supp_Mat_NR}.\eqref{eq:Onsager_Casimir_rel}], integration over volume $V$ of the surface $S$ defining the system (Fig.~\cite{Supp_Mat_NR}.\ref{fig:Multiport}) and application of Gauss theorem in LHS:
\begin{equation}\label{eq:vol_int_zero_RHS}
\oiint_S\left(\tilde{\ve{E}}'\times\tilde{\ve{H}}''
-\tilde{\ve{E}}''\times\ve{H}'\right)\cdot\uve{n}ds=0
\end{equation}
\noindent $\bullet$ Total field as the sum of fields at all the ports~\cite{Supp_Mat_NR}.\eqref{eq:WG_Modes}, with waveguide local reference planes placed at $z=0$ on the surface of the network:
\begin{subequations}\label{eq:WG_Modes_sum}
\begin{equation}
\tilde{\ve{E}}_{t}(x,y,z)
=\sum_p\left(a_p+b_p\right)\tilde{\ve{e}}_{t,p}(x,y)
\end{equation}
\begin{equation}
\tilde{\ve{H}}_{t}(x,y,z)
=\sum_p\left(a_p-b_p\right)\tilde{\ve{h}}_{t,p}(x,y)
\end{equation}
\end{subequations}
\noindent $\bullet$ Inserting single-primed and double-primed instances of~\eqref{eq:WG_Modes_sum} into~\eqref{eq:vol_int_zero_RHS} yields:
\begin{equation}\label{eq:IS1mS2}
\oiint_S\left(\tilde{\ve{E}}'\times\tilde{\ve{H}}''
-\tilde{\ve{E}}''\times\ve{H}'\right)\cdot\uve{n}ds=0
=I_1-I_2
\end{equation}
\begin{subequations}
\footnotesize
\begin{equation}
\begin{split}
I_1
&=\oiint_S\left[
\sum_p\left(a_p'+b_p'\right)\tilde{\ve{e}}_{t,p}'(x,y)
\times
\sum_q\left(a_q''-b_q''\right)\tilde{\ve{h}}_{t,q}''(x,y)
\right]\cdot\uve{n}ds \\
&=\sum_p\sum_q\left(a_p'a_q''-a_p'b_q''+b_p'a_q''-b_p'b_q''\right)
\underbrace{\oiint_S\left[\tilde{\ve{e}}_{t,p}'(x,y)
\times\tilde{\ve{h}}_{t,q}''(x,y)\right]\cdot\uve{n}ds}_{=2\delta_{pq}\tx{ by orthogonality (Sec.~\cite{Supp_Mat_NR}.\ref{sec:scat_pat_mod})}} \\
&=2\sum_p\left(a_p'a_p''-a_p'b_p''+b_p'a_p''-b_p'b_p''\right)
\end{split}
\end{equation}
\begin{equation}
\begin{split}
I_2
&=\oiint_S\left[
\sum_p\left(a_p''+b_p''\right)\tilde{\ve{e}}_{t,p}''(x,y)
\times
\sum_q\left(a_q'-b_q'\right)\tilde{\ve{h}}_{t,q}'(x,y)
\right]\cdot\uve{n}ds \\
&=\sum_p\sum_q\left(a_p''a_q'-a_p''b_q'+b_p''a_q'-b_p''b_q'\right)
\underbrace{\oiint_S\left[\tilde{\ve{e}}_{t,p}''(x,y)
\times\tilde{\ve{h}}_{t,q}'(x,y)\right]\cdot\uve{n}ds}_{=2\delta_{pq}\tx{ by orthogonality (Sec.~\cite{Supp_Mat_NR}.\ref{sec:scat_pat_mod})}} \\
&=2\sum_p\left(a_p''a_p'-a_p''b_p'+b_p''a_p'-b_p''b_p'\right)
\end{split}
\end{equation}
so that
\begin{equation}\label{eq:I1mI2}
\begin{split}
&I_1-I_2\\
&=2\sum_p\left(\cancel{a_p'a_p''}-a_p'b_p''+b_p'a_p''-\bcancel{b_p'b_p''}\right)
-\left(\cancel{a_p''a_p'}-a_p''b_p'+b_p''a_p'-\bcancel{b_p''b_p'}\right) \\
&=4\sum_p\left(b_p'a_p''-a_p'b_p''\right) \\
\end{split}
\end{equation}
\end{subequations}
\noindent $\bullet$ Inserting~\eqref{eq:I1mI2} into~\eqref{eq:IS1mS2}:
\begin{equation}\label{eq:ab_diff}
\sum_p\left(b_p'a_p''-a_p'b_p''\right)=0
\end{equation}
\noindent $\bullet$ Expanding the LHS of~\eqref{eq:ab_diff}:
\begin{equation}\label{eq:S_el_to_vec}
\begin{split}
&\sum_p\left(b_p'a_p''-a_p'b_p''\right)\\
&\;=[b_1',b_2',\ldots][a_1'',a_2'',\ldots]^T
-[a_1',a_2',\ldots][b_1'',b_2'',\ldots] \\
&\;=\ve{b}'\ve{a}^{\prime\prime T}-\ve{a}'\ve{b}^{\prime\prime T}=0
\end{split}
\end{equation}
\noindent $\bullet$ Recalling the definition of the S-matrix [Eq.~\cite{Supp_Mat_NR}.\eqref{eq:b_eq_Sa}]:
\begin{subequations}\label{eq:b_eq_Sa_A}
\begin{equation}
\ve{b}=\ve{S}\ve{a}
\end{equation}
\begin{equation}
\forall\quad\ve{b}=[b_1,b_2,\ldots]^T
\quad\tx{and}\quad
\ve{a}=[a_1,a_2,\ldots]^T
\end{equation}
\end{subequations}
\noindent $\bullet$ Using~\eqref{eq:b_eq_Sa_A} to eliminate $\ve{b}'$ and $\ve{b}''$ in~\eqref{eq:S_el_to_vec}:
\begin{equation}\label{eq:S_el_to_vec_aS}
\begin{split}
\ve{b}'\ve{a}^{\prime\prime T}-\ve{a}'\ve{b}^{\prime\prime T}
&=(\ve{S}\ve{a}')\ve{a}^{\prime\prime T}-\ve{a}'(\ve{S}\ve{a}'')^T \\
&=\ve{S}\ve{a}'\ve{a}^{\prime\prime T}-\ve{a}'\ve{a}^{\prime\prime T}\ve{S}^T \\
&=\ve{a}'\ve{a}^{\prime\prime T}\ve{S}-\ve{a}'\ve{a}^{\prime\prime T}\ve{S}^T \\
&=\ve{a}'\ve{a}^{\prime\prime T}(\ve{S}-\ve{S}^T) \\
&=0
\end{split}
\end{equation}

\noindent $\bullet$ Since this equation must hold true for any source sets $\ve{a}'$ and $\ve{a}''$, one must have:
\begin{equation}
\ve{S}-\ve{S}^T=0
\quad\tx{or}\quad
\ve{S}=\ve{S}^T
\end{equation}
$\equiv$ Eq.~\cite{Supp_Mat_NR}.\eqref{eq:Sp_NR}

\subsection{Inexistence of a Medium-Based Generalized Reciprocity Relation for an LTV Multiport Network}

\noindent$\rightarrow$ Eq.~\cite{Supp_Mat_NR}.\eqref{eq:gen_rec_the_for_S}

\noindent $\bullet$ Maxwell equations for set of prime\footnote{Warning: Here the prime does not denote TR but only another set of excitation-response fields.} (excitation-response) fields:
\begin{subequations}\label{eq:Maxwell_A2}
\begin{equation}\label{eq:Faraday_p}
\nabla\times\ve{E}'=-\partial\ve{B}'/\partial t
\end{equation}
\vspace{-5mm}
\begin{equation}\label{eq:Ampere_p}
\nabla\times\ve{H}'=\partial\ve{D}'/\partial t+\ve{J}'
\end{equation}
\end{subequations}

\noindent $\bullet$ Maxwell equations for set of double-prime (excitation-response) fields:
\begin{subequations}
\begin{equation}\label{eq:Faraday_dp}
\nabla\times\ve{E}''=-\partial\ve{B}''/\partial t
\end{equation}
\vspace{-5mm}
\begin{equation}\label{eq:Ampere_dp}
\nabla\times\ve{H}''=\partial\ve{D}''/\partial t+\ve{J}''
\end{equation}
\end{subequations}
\noindent $\bullet$ Subtracting~\eqref{eq:Faraday_p} dot multiplied by $\ve{H}''$ from~\eqref{eq:Ampere_dp}
dot multiplied by $\ve{E}'$, and doing the same with swapped prime and double-prime quantities:
\begin{subequations}
\begin{equation}\label{eq:Lmanipa2}
\ve{E}'\cdot\nabla\times\ve{H}''-\ve{H}''\cdot\nabla\times\ve{E}'
=\ve{E}'\cdot\partial\ve{D}''/\partial t+\ve{E}'\cdot\ve{J}''+\ve{H}''\cdot\partial\ve{B}'/\partial t
\end{equation}
\begin{equation}\label{eq:Lmanipb2}
\ve{E}''\cdot\nabla\times\ve{H}'-\ve{H}'\cdot\nabla\times\ve{E}''
=\ve{E}''\cdot\partial\ve{D}'/\partial t+\ve{E}''\cdot\ve{J}'+\ve{H}'\cdot\partial\ve{B}''/\partial t
\end{equation}
\end{subequations}
and, applying the identity $\ve{A}\cdot\nabla\times\ve{B}-\ve{B}\cdot\nabla\times\ve{A}=-\nabla\cdot(\ve{A}\times\ve{B})$:
\begin{subequations}\label{eq:Lmanip3}
\begin{equation}\label{eq:Lmanipa3}
-\nabla\cdot(\ve{E}'\times\ve{H}'')
=\ve{E}'\cdot\partial\ve{D}''/\partial t+\ve{E}'\cdot\ve{J}''+\ve{H}''\cdot\partial\ve{B}'/\partial t
\end{equation}
\begin{equation}\label{eq:Lmanipb3}
-\nabla\cdot(\ve{E}''\times\ve{H}')
=\ve{E}''\cdot\partial\ve{D}'/\partial t+\ve{E}''\cdot\ve{J}'+\ve{H}'\cdot\partial\ve{B}''/\partial t
\end{equation}
\end{subequations}

\noindent $\bullet$ Subtracting~\eqref{eq:Lmanipa3} from~\eqref{eq:Lmanipb3}:
\begin{equation}\label{eq:LcomplTD}
\begin{split}
&\nabla\cdot(\ve{E}'\times\ve{H}''-\ve{E}''\times\ve{H}')\\
&\quad=\ve{E}''\cdot\partial\ve{D}'/\partial t-\ve{E}'\cdot\partial\ve{D}''/\partial t \\
&\qquad+\ve{H}'\cdot\partial\ve{B}''/\partial t-\ve{H}''\cdot\partial\ve{B}'/\partial t \\
&\qquad+\ve{E}''\cdot\ve{J}'-\ve{E}'\cdot\ve{J}''
\end{split}
\end{equation}
\noindent $\bullet$ Eliminating the terms involving currents since the currents are assumed to be outside of the network (Fig.~\cite{Supp_Mat_NR}.\ref{fig:Multiport}) reduces~\eqref{eq:LcomplTD} to:
\begin{equation}\label{eq:LcomplTDwC}
\begin{split}
&\nabla\cdot\left(\ve{E}'\times\ve{H}''-\ve{E}''\times\ve{H}'\right)\\
&\quad=\left(\ve{E}''\cdot\partial\ve{D}'/\partial t-\ve{E}'\cdot\partial\ve{D}''/\partial t\right. \\
&\qquad\quad\left.+\ve{H}'\cdot\partial\ve{B}''/\partial t-\ve{H}''\cdot\partial\ve{B}'/\partial t\right)
\end{split}
\end{equation}
\noindent $\bullet$ Inserting the \emph{time-domain version} of the constitutive relations of~\eqref{eq:TR_bianisotropic_rel_dir} in order to accommodate LTV (linear time-variant) [$\dya{\epsilon}=\dya{\epsilon}(t)$, etc.] and NL (nonlinear) [$\dya{\epsilon}=\dya{\epsilon}(\ve{E},\ve{H})$, etc.] media,
\begin{subequations}\label{eq:TR_bianisotropic_rel_dir_TD}
\begin{equation}
\ve{D}
=\dya{\epsilon}(\ve{F}_0)*\ve{E}
+\dya{\xi}(\ve{F}_0)*\ve{H}
\end{equation}
\begin{equation}
\ve{B}
=\dya{\zeta}(\ve{F}_0)*\ve{E}
+\dya{\mu}(\ve{F}_0)*\ve{H}
\end{equation}
\end{subequations}
into~\eqref{eq:LcomplTDwC}:
\begin{equation}\label{eq:LcomplTDwC_const_par}
\begin{split}
&\nabla\cdot\left(\ve{E}'\times\ve{H}''-\ve{E}''\times\ve{H}'\right) \\
&\quad=\ve{E}''\cdot\partial\left[\dya{\epsilon}(\ve{F}_0)*\ve{E}'+\dya{\xi}(\ve{F}_0)*\ve{H}'\right]/\partial t \\
&\qquad\quad-\ve{E}'\cdot\partial\left[\dya{\epsilon}(\ve{F}_0)*\ve{E}''+\dya{\xi}(\ve{F}_0)*\ve{H}''\right]/\partial t \\
&\qquad\quad+\ve{H}'\cdot\partial\left[\dya{\zeta}(\ve{F}_0)*\ve{E}''+\dya{\mu}(\ve{F}_0)*\ve{H}''\right]/\partial t \\
&\qquad\quad-\ve{H}''\cdot\partial\left[\dya{\zeta}(\ve{F}_0)*\ve{E}'+\dya{\mu}(\ve{F}_0)*\ve{H}'\right]/\partial t
\end{split}
\end{equation}
or, applying the relation $\partial(f*g)/\partial t=\partial f/\partial t*g=f*\partial g/\partial t$:
\begin{equation}\label{eq:LcomplTDwC_const_par_conf_form}
\begin{split}
&\nabla\cdot\left[\ve{E}'\times\ve{H}''-\ve{E}''\times\ve{H}'\right] \\
&\quad=\ve{E}''\cdot\left(\dya{\epsilon}(\ve{F}_0)*\partial\ve{E}'/\partial t+\dya{\xi}(\ve{F}_0)*\partial \ve{H}'/\partial t\right) \\
&\qquad\quad-\ve{E}'\cdot\left[\dya{\epsilon}(\ve{F}_0)*\partial\ve{E}''/\partial t+\dya{\xi}(\ve{F}_0)*\partial\ve{H}''/\partial t\right] \\
&\qquad\quad+\ve{H}'\cdot\left[\dya{\zeta}(\ve{F}_0)*\partial\ve{E}''/\partial t+\dya{\mu}(\ve{F}_0)*\partial\ve{H}''/\partial t\right] \\
&\qquad\quad-\ve{H}''\cdot\left[\dya{\zeta}(\ve{F}_0)*\partial\ve{E}'/\partial t+\dya{\mu}(\ve{F}_0)*\partial\ve{H}'/\partial t\right]
\end{split}
\end{equation}
or
\begin{equation}\label{eq:LcomplTDwC_const_par_conf_nb}
\begin{split}
&\nabla\cdot\left(\ve{E}'\times\ve{H}''-\ve{E}''\times\ve{H}'\right) \\
&\quad=\ve{E}''\cdot\dya{\epsilon}(\ve{F}_0)*\partial\ve{E}'/\partial t+\ve{E}''\cdot\dya{\xi}(\ve{F}_0)*\partial \ve{H}'/\partial t
\\
&\qquad\quad-\ve{E}'\cdot\dya{\epsilon}(\ve{F}_0)*\partial\ve{E}''/\partial t-\ve{E}'\cdot\dya{\xi}(\ve{F}_0)*\partial\ve{H}''/\partial t \\
&\qquad\quad+\ve{H}'\cdot\dya{\zeta}(\ve{F}_0)*\partial\ve{E}''/\partial t+\ve{H}'\cdot\dya{\mu}(\ve{F}_0)*\partial\ve{H}''/\partial t \\
&\qquad\quad-\ve{H}''\cdot\dya{\zeta}(\ve{F}_0)*\partial\ve{E}'/\partial t-\ve{H}''\cdot\dya{\mu}(\ve{F}_0)*\partial\ve{H}'/\partial t
\end{split}
\end{equation}
and, grouping terms with same constitutive parameters:
\begin{equation}\label{eq:LcomplTDwC_const_par_conf_gr}
\begin{split}
&\nabla\cdot\left(\ve{E}'\times\ve{H}''-\ve{E}'\times\ve{H}'\right) \\
&\quad=\ve{E}''\cdot\dya{\epsilon}(\ve{F}_0)*\partial\ve{E}'/\partial t
-\ve{E}'\cdot\dya{\epsilon}(\ve{F}_0)*\partial\ve{E}''/\partial t\\
&\qquad\quad+\ve{H}'\cdot\dya{\mu}(\ve{F}_0)*\partial\ve{H}''/\partial t
-\ve{H}''\cdot\dya{\mu}(\ve{F}_0)*\partial\ve{H}'/\partial t \\
&\qquad\quad+\ve{E}''\cdot\dya{\xi}(\ve{F}_0)*\partial \ve{H}'/\partial t
-\ve{E}'\cdot\dya{\xi}(\ve{F}_0)*\partial\ve{H}''/\partial t \\
&\qquad\quad+\ve{H}'\cdot\dya{\zeta}(\ve{F}_0)*\partial\ve{E}''/\partial t
-\ve{H}''\cdot\dya{\zeta}(\ve{F}_0)*\partial\ve{E}'/\partial t
\end{split}
\end{equation}

\noindent $\bullet$ There is no way, even approximate, that Eq.~\eqref{eq:LcomplTDwC_const_par_conf_gr} can reduce to an equation of the type of Eq.~\eqref{eq:gen_rec_the_for_S_App} [or Eq.~\cite{Supp_Mat_NR}.\eqref{eq:gen_rec_the_for_S}]. So, this relation is used erroneously in~\cite{Jalas_2013} when referring to nonlinear systems.

\end{document}